\documentclass[sigplan,10pt]{acmart}

\newif\iflong
\longtrue %

\usepackage{xspace}
\usepackage{dsfont}
\usepackage{macros}

\usepackage[noline,noend,ruled,linesnumbered]{algorithm2e}
\usepackage{algo}

\usepackage{etoolbox}

\makeatletter
\patchcmd{\@algocf@start}%
  {-1.5em}%
  {0pt}%
  {}{}%
\makeatother

\usepackage[all]{xy}
\usepackage{multirow}

\usepackage{subcaption}

\usepackage{placeins}

\usepackage{pifont}

\newcommand{\rmark}{\ding{51}\xspace}
\newcommand{\wmark}{\ding{55}\xspace}

\usepackage{tikz}

\usepackage{paralist}

\usepackage{multicol}

\usepackage{lipsum}

\usepackage{ifsym}

\newcommand\phd[1]{}

\copyrightyear{2021}
\acmYear{2021}
\setcopyright{acmcopyright}\acmConference[EuroSys '21]{Sixteenth European Conference on Computer Systems}{April 26--28, 2021}{Online, United Kingdom}
\acmBooktitle{Sixteenth European Conference on Computer Systems (EuroSys '21), April 26--28, 2021, Online, United Kingdom}
\acmPrice{15.00}
\acmDOI{10.1145/3447786.3456236}
\acmISBN{978-1-4503-8334-9/21/04}

\begin{document}

\title{Efficient Replication via Timestamp Stability}

\author{Vitor Enes}
\affiliation{%
  \institution{INESC TEC and\\University of Minho}
}

\author{Carlos Baquero}
\affiliation{%
  \institution{INESC TEC and\\University of Minho}
}

\author{Alexey Gotsman}
\affiliation{%
  \institution{IMDEA Software Institute}
}

\author{Pierre Sutra}
\affiliation{%
  \institution{T\'el\'ecom SudParis}
}



\begin{abstract}
  Modern web applications replicate their data across the globe and require strong consistency guarantees for their most critical data. 
  These guarantees are usually provided via state-machine replication (SMR).
  Recent advances in SMR have focused on leaderless protocols, which improve the availability and performance of traditional Paxos-based solutions.
  We propose \SYS{} -- a leaderless SMR protocol that, in comparison to prior solutions, achieves superior throughput and offers predictable performance even in contended workloads.
  To achieve these benefits, \SYS{} {\em timestamps} each application command and executes it only after the timestamp becomes {\em stable}, i.e., all commands with a lower timestamp are known.
  Both the timestamping and stability detection mechanisms are fully decentralized, thus obviating the need for a leader replica.
  Our protocol furthermore generalizes to partial replication settings, enabling scalability in highly parallel workloads.
  We evaluate the protocol in both real and simulated geo-distributed environments and demonstrate that it outperforms state-of-the-art alternatives.
\end{abstract}

\maketitle

\section{Introduction}
\label{sec:introduction}

Modern web applications are routinely accessed by clients all over the world.
To support such applications, storage systems need to replicate data at different geographical locations while providing strong consistency guarantees for the most critical data.
\emph{State-machine replication (SMR)}~\cite{smr} is an approach for providing such guarantees used by a number of systems \cite{spanner,cassandra,cockroachdb,yugabytedb,faunadb,azure}.
In SMR, a desired service is defined by a deterministic state machine, and each site maintains its own local replica of the machine.
An {\em SMR protocol} coordinates the execution of commands at the sites to ensure that the system is {\em linearizable}~\cite{linearizability}, i.e., behaves as if commands are executed sequentially by a single site.

Traditional SMR protocols, such as Paxos~\cite{paxos} and Raft~\cite{raft}, rely on a distinguished {\em leader} site that defines the order in which client commands are executed at the replicas.
Unfortunately, this site is a single point of failure and contention, and a source of higher latency for clients located far from it.
Recent efforts to improve SMR have thus focused on {\em leaderless} protocols, which distribute the task of ordering commands among replicas and thus allow a client to contact the closest replica instead of the leader~\cite{mencius,atlas,gryff,caesar,epaxos,alvin}.
Compared to centralized solutions, leaderless SMR offers lower average latency, fairer latency distribution with respect to client locations, and higher availability.

Leaderless SMR protocols also generalize to the setting of {\em partial replication}, where the service state is split into a set of partitions, each stored at a group of replicas.
A client command can access multiple partitions, and the SMR protocol ensures that the system is still linearizable, i.e., behaves as if the commands are executed by a single machine storing a complete service state.
This approach allows implementing services that are too big to fit onto a single machine.
It also enables scalability, since commands accessing disjoint sets of partitions can be executed in parallel.
This has been demonstrated by Janus~\cite{janus} which adapted a leaderless SMR protocol called Egalitarian Paxos (EPaxos)~\cite{epaxos} to the setting of partial replication.
The resulting protocol provided better performance than classical solutions such as two-phase commit layered over Paxos.

Unfortunately, all existing leaderless SMR protocols suffer from drawbacks in the way they order commands.
Some protocols~\cite{epaxos,atlas,caesar,gryff} maintain explicit dependencies between commands: a replica may execute a command only after all its dependencies get executed.
These dependencies may form arbitrary long chains.
As a consequence, in theory the protocols do not guarantee progress even under a synchronous network.
In practice, their performance is unpredictable, and in particular, exhibits a high tail latency~\cite{gryff,leaderless}.
Other protocols~\cite{mencius,clockrsm} need to contact every replica on the critical path of each command.
While these protocols guarantee progress under synchrony, they make the system run at the speed of the slowest replica.

All of these drawbacks carry over to the setting of partial replication where they are aggravated by the fact that commands span multiple machines.

In this paper we propose \SYS, a new leaderless SMR protocol that lifts the above limitations while handling both full and partial replication settings.
\SYS guarantees progress under a synchronous network without the need to contact all replicas.
It also exhibits low tail latency even in contended workloads, thus ensuring predictable performance.
Finally, it delivers superior throughput than prior solutions, such as EPaxos and Janus.
The protocol achieves all these benefits by assigning a scalar {\em timestamp} to each command and executing commands in the order of these timestamps.
To determine when a command can be executed, each replica waits until the command's timestamp is {\em stable}, i.e., all commands with a lower timestamp are known.
Ordering commands in this way is used in many protocols~\cite{causality,granola,clockrsm,caesar}.
A key novelty of \SYS is that both timestamping and stability detection are fault-tolerant and fully decentralized, which preserves the key benefits of leaderless SMR.

In more detail, each \SYS process maintains a local clock from which timestamps are generated.
In the case of full replication, to submit a command a client sends it to the closest process, which acts as its {\em coordinator}.
The coordinator computes a timestamp for the command by forwarding it to a quorum of replicas, each of which makes a {\em timestamp proposal}, and taking the maximum of these proposals.
If enough replicas in the quorum make the same proposal, then the timestamp is decided immediately ({\em fast path}).
If not, the coordinator does an additional round trip to the replicas to persist the timestamp ({\em slow path}); this may happen when commands are submitted concurrently.
Thus, under favorable conditions, the replica nearest to the client decides the command's timestamp in a single round trip.

To execute a command, a replica then needs to determine when its timestamp is stable, i.e., it knows about all commands with lower timestamps.
The replica does this by gathering information about which timestamp ranges have been used up by each replica, so that no more commands will get proposals in these ranges.
This information is piggy-backed on replicas' messages, which often allows a timestamp of a command to become stable immediately after it is decided.

The above protocol easily extends to partial replication: in this case a command's timestamp is the maximum over the timestamps computed for each of the partitions it accesses.

We evaluate \SYS in three environments: a simulator, a controlled cluster environment and using multiple regions in Amazon EC2.
We show that \SYS improves throughput over existing SMR protocols by \textsf{1.8-5.1x}, while lowering tail latency with respect to prior leaderless protocols by an order of magnitude.
This advantage is maintained in partial replication, where \SYS outperforms Janus by \textsf{1.2-16x}.

\section{Partial State-Machine Replication}
\label{sec:model}

We consider a geo-distributed system where processes may fail by crashing, but do not behave maliciously.
State-machine replication (SMR) is a common way of implementing fault-tolerant services in such a system~\cite{smr}.
In SMR, the service is defined as a deterministic state machine accepting a set of {\em commands} $\cmds$.
Each process maintains a replica of the machine and receives commands from clients, external to the system.
An SMR protocol coordinates the execution of commands at the processes, ensuring that they stay in sync.

We consider a general version of SMR where each process replicates only a part of the service state -- {\em partial SMR (PSMR)}~\cite{epidemic,pstore,janus}.
We assume that the service state is divided into \emph{partitions}, so that each variable defining the state belongs to a unique partition.
Partitions are arbitrarily fine-grained: e.g., just a single state variable.
Each command accesses one or more partitions.
We assume that a process replicates a single partition, but multiple processes may be co-located at the same machine.
Each partition is replicated at $r$ processes, of which at most $f$ may fail.
Following Flexible Paxos~\cite{flexible-paxos}, $f$ can be any value such that $1 \le f \le \lfloor \frac{r-1}{2} \rfloor$.
This allows using small values of $f$ regardless of the replication factor $r$, which is appropriate in geo-replication~\cite{spanner,atlas}.
We write $\procs_p$ for the set of all the processes replicating a partition $p$, $\procs_c$ for the set of processes that replicate the partitions accessed by a command $c$, and $\procs$ for the set of all processes.

A PSMR protocol allows a process $i$ to submit a command $c$ on behalf of a client.
For simplicity, we assume that each command is unique and the process submitting it replicates one of the partitions it accesses: $i \in \procs_c$.
For each partition $p$ accessed by $c$, the protocol then triggers an upcall $\exec_p(c)$ at each process storing $p$, asking it to apply $c$ to the local state of partition $p$.
After $c$ is executed by at least one process in each partition it accesses, the process that submitted the
command aggregates the return values of $c$ from each partition and returns them to the client.

PSMR ensures the highest standard of consistency of replicated data -- {\em linearizability}~\cite{linearizability} -- which provides an illusion that commands are executed sequentially by a single machine storing a complete service state.
To this end, a PSMR protocol has to satisfy the following specification.
Given two commands $c$ and $d$, we write $c \order_i d$ when they access a common partition and $c$ is executed before $d$ at some process $i \in \procs_c \inter \procs_d$.
We also define the following {\em real-time order}: $c \rt d$ when the command $c$ returns before the command $d$ was submitted.
Let $\order~= (\bigcup_{i \in \procs} {\order_i}) \union {\rt}$.
A PSMR protocol ensures the following properties:

\vspace{0.5em}

\textbf{Validity.}
If a process executes some command $c$, then it executes $c$ at most once and only if $c$ was submitted before.

\textbf{Ordering.}
The relation $\order$ is acyclic.

\textbf{Liveness.}
If a command $c$ is submitted by a non-faulty process or executed at some process, then it is executed at all non-faulty processes in $\procs_c$.

\vspace{0.5em}

The Ordering property ensures that commands are executed in a consistent manner throughout the system \cite{zoofence}.
For example, it implies that two commands, both accessing the same two partitions, cannot be executed at these partitions in contradictory orders.
As usual, to ensure Liveness we assume that the network is eventually synchronous, and in particular, that message delays between non-failed processes are eventually bounded~\cite{psync}.

PSMR is expressive enough to implement a wide spectrum of distributed applications.
In particular, it directly allows implementing one-shot transactions, which consist of independent pieces of code (such as stored procedures), each accessing a different partition~\cite{hstore,janus,snow}.
It can also be used to construct general-purpose transactions~\cite{determinism,janus}. %

\section{Single-Partition Protocol}
\label{sec:sys}

For simplicity, we first present the protocol in the case when there is
only a single partition, and cover the general case in \refsec{sec:sys_partial}.
We start with an overview of the single-partition protocol.

To ensure the Ordering property of PSMR, \SYS assigns a scalar {\em timestamp} to each command.
Processes execute commands in the order of these timestamps, thus ensuring that processes execute commands in the same order.
To submit a command, a client sends it to a nearby process which acts as the {\em coordinator} for the command.
The coordinator is in charge of assigning a timestamp to the command and communicating this timestamp to all processes.
When a process finds out about the command's timestamp, we say that the process {\em commits} the command.
If the coordinator is suspected to have failed, another process takes over its role through a recovery mechanism (\refsec{sec:sys_recovery}).
\SYS ensures that, even in case of failures, processes agree on the timestamp assigned to the command, as stated by the following property.
\begin{property}[Timestamp agreement]
  \label{prop:agreement}
  \upshape
  Two processes cannot commit the same command with different timestamps.
\end{property}

A coordinator computes a timestamp for a command as follows (\refsec{sec:sys_commit}).
It first forwards the command to a {\em fast quorum} of $\floor{\frac{r}{2}}+f$ processes, including the coordinator itself.
Each process maintains a $\Clock$ variable.
When the process receives a command from the coordinator, it increments $\Clock$ and replies to the coordinator with the new $\Clock$ value as a \emph{timestamp proposal}.
The coordinator then takes the highest proposal as the command's timestamp.
If enough processes have made such a proposal, the coordinator considers the timestamp decided and takes the {\em fast path}: it just communicates the timestamp to the processes, which commit the command.
The protocol ensures that the timestamp can be recovered even if the coordinator fails, thus maintaining \refprop{prop:agreement}.
Otherwise, the coordinator takes the {\em slow path}, where it stores the timestamp at a {\em slow quorum} of $f+1$ processes using a variant of Flexible Paxos~\cite{flexible-paxos}.
This ensures that the timestamp survives any allowed number of failures.
The slow path may have to be taken in cases when commands are submitted concurrently to the same partition (however, recall that partitions may be arbitrarily fine-grained).

Since processes execute committed commands in the timestamp order, before executing a command a process must know all the commands that precede it. %
\begin{property}[Timestamp stability]
  \label{prop:stability}
  \upshape
  Consider a command $c$ committed at $i$ with timestamp $t$.
  Process $i$ can only execute $c$ after its timestamp is {\em stable}, i.e., every command with a timestamp lower or equal to $t$ is also committed at $i$.
\end{property}

To check the stability of a timestamp $t$ (\refsec{sec:sys_execution}), each process $i$ tracks timestamp proposals issued by other processes.
Once the $\Clock$s at any majority of the processes pass $t$, process $i$ can be sure that new commands will get higher timestamps: these are computed as the maximal proposal from at least a majority, and any two majorities intersect.
Process $i$ can then use the information gathered about the timestamp proposals from other processes to find out about all the commands that have got a timestamp lower than $t$.

\subsection{Commit Protocol}
\label{sec:sys_commit}

\begin{algorithm}
    \small

    \setcounter{AlgoLine}{0}

    \SubAlgo{$\FunSubmit(c)$ \label{algo:commit:submit}}{
        \Pre $i \in \procs_c$ \label{algo:commit:submit-pre} \;
        $\id \leftarrow \af{next\_id}()$;
        $\qs \leftarrow \af{fast\_quorums}(i, \procs_c)$ \;
        \Send $\MSubmit(\id, c, \qs)$ \To $\procs_c^i$
    }

    \SubAlgo{\Receive $\MSubmit(\id, c, \qs)$ \label{algo:commit:msubmit-recv}}{
        $t \leftarrow \Clock + 1$ \label{algo:commit:msubmit-proposal} \;
        \Send $\MPropose(\id, c, \qs, t)$ \To $\qs[p]$ \label{algo:commit:mpropose-send} \;
        \Send $\MPayload(\id, c, \qs)$ \To $\procs_p \setminus \qs[p]$ \label{algo:commit:mpayload-send} \;
    }

    \SubAlgo{\Receive $\MPayload(\id, c, \qs)$\label{algo:commit:mpayload-recv}}{
        \Pre $\id \in \start$ \label{algo:commit:mpayload-pre} \;
        \mbox{$\cmd[\id] \leftarrow c$;
        $\quorums[\id] \leftarrow \qs$;
        $\phase[\id] \leftarrow \PAYLOAD$ \label{algo:commit:payload-phase}}
    }

    \SubAlgo{\Receive $\MPropose(\id, c, \qs, t)$ \From $j$ \label{algo:commit:mpropose-recv}}{
        \Pre $\id \in \start$ \label{algo:commit:mpropose-pre} \;
        \mbox{$\cmd[\id] \leftarrow c$;
        $\quorums[\id] \leftarrow \qs$;
        $\phase[\id] \leftarrow \PROPOSE$ \label{algo:commit:propose-phase}} \;
        $\ts[\id] \leftarrow \FunTs(\id, t)$ \label{algo:commit:mpropose-proposal} \;
        \Send $\MProposeAck(\id, \ts[\id])$ \To $j$ \label{algo:commit:mpropose-ack-send} \;
    }

    \SubAlgo{\Receive $\MProposeAck(\id, t_j)$ \From $\forall j \in Q$ \label{algo:commit:mpropose-ack-recv}}{
        \Pre $\id \in \propose \land Q = \quorums[\id][p]$ \label{algo:commit:mpropose-ack-pre} \;
        $t \leftarrow \max \{ t_j \mid j \in Q \}$ \label{algo:commit:mpropose-ack-max} \;
        \lIf{$\af{count}(t) \geq f$\label{algo:commit:fast-path-condition}}{%
            \Send $\MCommit(\id, t)$ \To $\procs_{\cmd[\id]}$ \label{algo:commit:fast-path}
        }
        \lElse{%
            \Send $\MConsensus(\id, t, i)$ \To $\procs_p$ \label{algo:commit:slow-path}
        }
    }

    \SubAlgo{\Receive $\MCommit(\id, t)$\label{algo:commit:mcommit-recv}}{
        \Pre $\id \in \pending$ \label{algo:commit:mcommit-pre} \;
        $\ts[\id] \leftarrow t$;
        $\phase[\id] \leftarrow \COMMIT$ \label{algo:commit:commit-phase} \;
        $\FunBump(\ts[\id])$ \label{algo:commit:mcommit-bump}
    }

    \SubAlgo{\Receive $\MConsensus(\id, t, b)$ \From $j$}{
        \Pre $\bal[\id] \leq b$ \label{algo:commit:mconsensus-pre} \;
        $\ts[\id] \leftarrow t$;
        $\bal[\id] \leftarrow b$;
        $\abal[\id] \leftarrow b$ \;
        $\FunBump(t)$ \label{algo:commit:mconsensus-bump} \;
        \Send $\MConsensusAck(\id, b)$ \To $j$
    }

    \SubAlgo{\Receive $\MConsensusAck(\id, b)$ \From $Q$\label{algo:commit:receive-mconsensus}}{
        \Pre $\bal[\id] = b \land \setsize{Q} = f + 1$ \label{algo:commit:mconsensus-ack-pre} \;
        \Send $\MCommit(\id, \ts[\id])$ \To $\procs_{\cmd[\id]}$ \label{algo:commit:consensus-end}
    }
    \algrule[0.5pt]

    \SubAlgo{$\FunTs(\id, m)$ \label{algo:commit:fun-ts}}{
        $t \leftarrow \max(m, \Clock + 1)$ \label{algo:commit:timestamp-max} \;
        \mbox{$\Detached \leftarrow \Detached \,{\union}\, \{ \tup{i, u} \,{\mid}\, \Clock \,{+}\, 1 \,{\leq}\, u \,{\leq}\, t\,{-}\,1 \}$}\label{algo:commit:detached} \\
        $\Attached[\id] \leftarrow \{ \tup{i, t} \}$ \label{algo:commit:attached} \;
        $\Clock \leftarrow t$ \label{algo:commit:fun-ts-bump} \;
        \Return $t$ \label{algo:commit:fun-ts-return}
    }

    \SubAlgo{$\FunBump(t)$ \label{algo:commit:fun-bump}}{
        $t \leftarrow \max(t, \Clock)$ \;
        \mbox{$\Detached \leftarrow \Detached \union \{ \tup{i, u} \mid \Clock \,{+}\, 1 \leq u \leq t \}$} \\
        $\Clock \leftarrow t$
    }

    \newcounter{CommitLastLine}
    \setcounter{CommitLastLine}{\value{AlgoLine}}

    \caption{Commit protocol at process $i \in \procs_p$.}
    \label{algo:commit}
\end{algorithm}

\refalgo{algo:commit} specifies the single-partition commit protocol at a process $i$ replicating a partition $p$.
We assume that self-addressed messages are delivered immediately.
A command $c \in \cmds$ is submitted by a client by calling $\submit(c)$ at a process $i$ that replicates a partition accessed by the command (\refline{algo:commit:submit}).
Process $i$ then creates a unique identifier $\id \in \ids$ and a mapping $\qs$ from a partition accessed by the command to the fast quorum to be used at that partition.
Because we consider a single partition for now, in what follows $\qs$ contains only one fast quorum, $\qs[p]$.
Finally, process $i$ sends $\MSubmit(\id, c, \qs)$ to a set of processes $\procs_c^i$, which in the single-partition case simply denotes $\{i\}$.

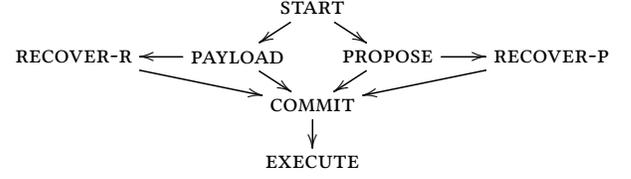
\begin{figure}[t]
    \begin{center}
        \begin{tabular}{c}
            \begin{xy}
                <0.5cm,0pt>:
                (0,3.3)*+{\START}="start";
                (-2,2)*+{\PAYLOAD}="payload";
                (-6.35,2)*+{\RECOVERR}="recoverr";
                (2,2)*+{\PROPOSE}="propose";
                (6.35,2)*+{\RECOVERP}="recoverp";
                (0,0.7)*+{\COMMIT}="commit";
                (0,-0.8)*+{\EXECUTE}="execute";
                \ar@{->} "propose"; "start";
                \ar@{->} "payload"; "start";
                \ar@{->} "recoverr"; "payload";
                \ar@{->} "recoverp"; "propose";
                \ar@{->} "commit"; "payload";
                \ar@{->} "commit"; "propose";
                \ar@{->} "commit"; "recoverr";
                \ar@{->} "commit"; "recoverp";
                \ar@{->} "execute"; "commit";
            \end{xy}
        \end{tabular}
    \end{center}

    \vspace{-0.45cm}
    \caption{
        Command journey through phases in \SYS.
    }
    \Description{TODO}
    \label{fig:phases}
\end{figure}

A command goes through several {\em phases} at
each process: from the initial phase $\START$, to a $\COMMIT$ phase once the
command is committed, and an $\EXECUTE$ phase once it is executed.
We summarize these phases and allowed phase transitions in \reffig{fig:phases}.
A mapping $\phase$ at a process tracks the progress of a command
with a given identifier
through phases. For brevity, the name of the phase written in lower case denotes
all the commands in that phase, e.g.,
$\start = \{\id \in \ids \mid \phase[\id] = \START \}$.
We also define $\pending$ as follows: $\pending = \payload \union \propose \union \recoverp \union \recoverr$.

\subsubsection*{Start phase}
When a process receives an $\MSubmit$ message, it starts serving as the command coordinator (\refline{algo:commit:msubmit-recv}).
The coordinator first computes its timestamp proposal
for the command as $\Clock + 1$.
After computing the proposal, the coordinator sends an $\MPropose$ message to the fast quorum $\qs[p]$ and an $\MPayload$ message to the remaining processes.
Since the fast quorum contains the coordinator, the coordinator also sends the $\MPropose$ message to itself. As mentioned earlier, self-addressed messages are delivered immediately.

\subsubsection*{Payload phase}
Upon receiving an $\MPayload$ message (\refline{algo:commit:mpayload-recv}), a process simply saves the command payload in a mapping $\cmd$ and sets the command's phase to $\PAYLOAD$.
It also saves $\qs$ in a mapping $\quorums$. This is necessary
for the recovery mechanism to know the fast quorum used for the command (\refsec{sec:sys_recovery}).

\subsubsection*{Propose phase}
Upon receiving an $\MPropose$ message (\refline{algo:commit:mpropose-recv}), a
fast-quorum process also saves the command payload and fast quorums, but sets its phase to $\PROPOSE$.
Then
the process computes its own timestamp proposal using the function $\FunTs$ and
stores it in a mapping $\ts$. Finally, the process replies to the coordinator
with an $\MProposeAck$ message, carrying the computed timestamp proposal.

The function $\FunTs$ takes as input an identifier $\id$ and a timestamp $m$
and computes a timestamp proposal as $t = \max(m, \Clock + 1)$,
so that $t \ge m$ (\refline{algo:commit:timestamp-max}).
The function \emph{bumps} the $\Clock$ to the computed timestamp $t$
and returns $t$ (lines~\ref{algo:commit:fun-ts-bump}-\ref{algo:commit:fun-ts-return}); we explain
lines~\ref{algo:commit:detached}-\ref{algo:commit:attached} later.
As we have already noted, the coordinator computes the command's timestamp as the
highest of the proposals from fast-quorum processes.
Proactively taking the $\max$ between the coordinator's proposal $m$ and
$\Clock + 1$ in $\FunTs$ ensures that a process's proposal is at least as
high as the coordinator's; as we explain shortly, this helps recovering
timestamps in case of coordinator failure.

\subsubsection*{Commit phase}
Once the coordinator receives an $\MProposeAck$ message from all the processes in
the fast quorum $Q = \qs[p]$ (\refline{algo:commit:mpropose-ack-recv}), it computes the
command's timestamp as the highest of all timestamp proposals:
$t = \max\{ t_j \mid j \in Q \}$.  Then the coordinator decides to either take
the fast path (\refline{algo:commit:fast-path-condition}) or the
slow path (\refline{algo:commit:slow-path}). Both paths end
with the coordinator sending an $\MCommit$ message containing the command's
timestamp. Since $\setsize{Q} = \floor{\frac{r}{2}} + f$ and $f \geq 1$, we have
the following property which ensures that a committed timestamp is computed over
(at least) a majority of processes.
\begin{property}
  \label{prop:bump-at-majority}
  \upshape
  For any message $\MCommit(\id, t)$, there is a set of processes $Q$ such that
  $|Q| \geq \floor{\frac{r}{2}} + 1$ and
  $t = \max\{ t_j \mid j \in Q \}$, where $t_j$ is the output of function
  $\FunTs(\id, \_)$ previously called at process $j \in Q$.
\end{property}
This property is also preserved if $t$ is computed by a process performing
recovery in case of coordinator failure (\refsec{sec:sys_recovery}).

Once a process receives an $\MCommit$ message
(\refline{algo:commit:mcommit-recv}), it saves the command's timestamp in
$\ts[\id]$ and moves the command to the $\COMMIT$ phase. It then bumps the
$\Clock$ to the committed timestamp using a function $\FunBump$
(\refline{algo:commit:fun-bump}).
We next explain the fast and slow paths, as well as the conditions under which
they are taken.

\subsubsection*{Fast path}
The fast path can be taken if the highest proposal $t$ is made by at least
$f$ processes.
This condition is expressed by $\af{count}(t) \geq f$ in
\refline{algo:commit:fast-path-condition}, where
$\af{count}(t) = \setsize{\{ j \in Q \mid t_j = t \}}$.
If the condition holds, the coordinator immediately sends an $\MCommit$ message
with the computed timestamp\footnote{%
  In \refline{algo:commit:fast-path} we send the message to $\procs_c$ even
  though this set is equal to $\procs_p$ in the single-partition case.
  We do this to reuse the pseudocode when presenting the multi-partition protocol
  in \refsec{sec:sys_partial}.
}.
The protocol ensures that, if the coordinator fails before sending all the $\MCommit$ messages, $t$ can be recovered as follows.
First, the condition $\af{count}(t) \geq f$ ensures that
the timestamp $t$ can be obtained without $f - 1$ fast-quorum processes
(e.g., if they fail) by selecting the highest proposal made by the remaining
quorum members.
Moreover, the proposal by the coordinator is also not necessary to obtain $t$.
This is because fast-quorum processes only propose timestamps no lower than
the coordinator's proposal (\refline{algo:commit:mpropose-proposal}).
As a consequence, the coordinator's proposal is only the highest proposal $t$
when all processes propose the same timestamp, in which case a single process
suffices to recover $t$.
It follows that $t$ can be obtained without $f$ fast-quorum processes
including the initial coordinator by selecting the
highest proposal sent by the remaining
$(\floor{\frac{r}{2}} + f) - f = \floor{\frac{r}{2}}$ quorum members.
This observation is captured by the following property.

\begin{property}
  \label{prop:recovery}
  \upshape
  Any timestamp committed on the fast path can be obtained by selecting the
  highest proposal sent in $\MPropose$ by at least $\floor{\frac{r}{2}}$
  fast-quorum processes distinct from the initial coordinator.
\end{property}

\subsubsection*{Fast path examples.}
\reftab{tab:fast_paths} contains several examples that illustrate
the fast-path condition of \SYS and \refprop{prop:recovery}.
All examples consider $r = 5$ processes. %
We highlight timestamp proposals in bold.
Process \A acts as the coordinator and sends
$\clockval{6}$ in its $\MPropose$ message.
The fast quorum $Q$ is $\{ \A, \B, \C \}$ when $f = 1$ and
$\{ \A, \B, \C, \D \}$ when $f = 2$.
The example in \reftab{tab:fast_paths} $a)$ considers \SYS $f = 2$.
Once process \B receives the $\MPropose$ with timestamp $\clockval{6}$, it bumps
its $\Clock$ from 6 to $\clockval{7}$ and sends a proposal $\clockval{7}$ in
the $\MProposeAck$.
Similarly, processes \C and \D bump their $\Clock$ from 10 to $\clockval{11}$
and propose $\clockval{11}$.
Thus, \A receives proposals
$t_\A = \clockval{6}$,
$t_\B = \clockval{7}$,
$t_\C = \clockval{11}$ and
$t_\D = \clockval{11}$, and
computes the command's timestamp as
$t = \max\{6, 7, 11\} = 11$.
Since $\af{count}(11) = 2 \geq f$, the coordinator takes the fast path, even though
the proposals did not match.
In order to understand why this is safe, assume that the coordinator fails (before sending all the $\MCommit$ messages) along with another fast-quorum process.
Independently of which $\floor{\frac{r}{2}} = 2$ fast-quorum processes survive
($\{ \B, \C \}$ or $\{ \B, \D \}$ or $\{ \C, \D \}$), timestamp $11$ is
always present and can be recovered as stated by \refprop{prop:recovery}.
This is not the case for the example in \reftab{tab:fast_paths} $b)$.
Here \A receives
$t_\A = \clockval{6}$,
$t_\B = \clockval{7}$,
$t_\C = \clockval{11}$ and
$t_\D = \clockval{6}$, and again computes
$t = \max\{6, 7, 11\} = 11$.
Since $\af{count}(11) = 1 < f$, the coordinator cannot take the fast path:
timestamp $11$ was proposed solely by \C and would be lost if both this process and the coordinator fail.
The examples in \reftab{tab:fast_paths} $c)$ and $d)$ consider $f = 1$, and the
fast path is taken in both, independently of the timestamps proposed.
This is because \SYS fast-path condition
$\af{count}(\max\{ t_j \mid j \in Q \}) \geq f$ trivially holds with $f = 1$,
and thus \SYS $f = 1$ always takes the fast path.

Note that when the $\Clock$ at a fast-quorum process is below the proposal
$m$ sent by the coordinator, i.e., $\Clock < m$, the process makes the same proposal as the coordinator.
This is not the case when $\Clock \geq m$, which can happen when
commands are submitted concurrently to the partition.
Nonetheless, \SYS is able to take the fast path in some of these
situations, as illustrated in \reftab{tab:fast_paths}.

\begin{table}
    \caption{
        \SYS examples with $r = 5$ processes while tolerating $f$ faults.
        Only 4 processes are depicted, \A, \B, \C and \D, with \A always acting
        as the coordinator.
    }
    \label{tab:fast_paths}
    \vspace{-0.30cm}

    \setlength\tabcolsep{2.5pt}
    \setlength\aboverulesep{0pt}
    \setlength\belowrulesep{0pt}

    \begin{center}
        \begin{tabular}{c|c|c|c|c|c|c}
                         & \A           & \B                    & \C                      & \D                      & match  & fast path
            \\
            \toprule
            $a)$ $f = 2$ & \clockval{6} & \clockevolution{6}{7} & \clockevolution{10}{11} & \clockevolution{10}{11} & \wmark & \rmark
            \\
            $b)$ $f = 2$ & \clockval{6} & \clockevolution{6}{7} & \clockevolution{10}{11} & \clockevolution{5}{6}   & \wmark & \wmark
            \\
            $c)$ $f = 1$ & \clockval{6} & \clockevolution{6}{7} & \clockevolution{10}{11} &                         & \wmark & \rmark
            \\
            $d)$ $f = 1$ & \clockval{6} & \clockevolution{5}{6} & \clockevolution{1}{6}   &                         & \rmark & \rmark
            \\
            \toprule
        \end{tabular}
    \end{center}
    \vspace{-0.30cm}
\end{table}

\subsubsection*{Slow path.}
When the fast-path condition does not hold, the timestamp
computed by the coordinator is not yet guaranteed to be persistent:
if the coordinator fails before sending all the $\MCommit$ messages, a process taking over its job may compute a different
timestamp. To maintain \refprop{prop:agreement} in this case,
the coordinator first reaches an agreement on the computed timestamp with other processes replicating the same partition.
This is implemented using single-decree Flexible Paxos~\cite{flexible-paxos}.
For each identifier we allocate ballot numbers to processes round-robin, with
ballot $i$ reserved for the initial coordinator $i$ and ballots higher than $r$
for processes performing recovery.
Every process stores for each identifier $\id$ the ballot $\bal[\id]$ it is
currently participating in and the last ballot $\abal[\id]$ in which it
accepted a consensus proposal (if any).
When the initial coordinator $i$ decides to go onto the slow path, it performs
an analog of Paxos Phase 2: it sends an $\MConsensus$ message with its
consensus proposal and ballot $i$ to a {\em slow quorum} that includes itself.
Following Flexible Paxos, the size of the slow quorum is
only $f + 1$, rather than a majority like in classical Paxos.
As usual in Paxos, a process accepts an $\MConsensus$ message only if its
$\bal[\id]$ is not greater than the ballot in the message
(\refline{algo:commit:mconsensus-pre}).
Then it stores the consensus proposal, sets $\bal[\id]$ and $\abal[\id]$ to the
ballot in the message, and replies to the coordinator with $\MConsensusAck$.
Once the coordinator gathers $f + 1$ such replies
(\refline{algo:commit:receive-mconsensus}), it is sure that its consensus
proposal will survive the allowed number of failures $f$, and it thus broadcasts
the proposal in an $\MCommit$ message.

\subsection{Execution Protocol}
\label{sec:sys_execution}

A process executes committed commands in the timestamp order.
To this end, as required by \refprop{prop:stability}, a process executes a command only after its timestamp becomes stable, i.e., all commands with a lower timestamp are known.
To detect stability, \SYS tracks which timestamp ranges have been used up by each process using the following mechanism.

\subsubsection*{Promise collection}
A \emph{promise} is a pair $\tup{j, u} \subseteq \procs_p \times \nat$ where $j$ is a process and $u$ a timestamp.
Promises can be \emph{attached} to some command or \emph{detached}.
A promise $\tup{j, u}$ attached to command $c$ means that process $j$ proposed timestamp $u$ for command $c$, and thus will not use this timestamp again.
A detached promise $\tup{j, u}$ means that process $j$ will never propose timestamp $u$ for any command.

The function $\FunTs$ is responsible for collecting the promises issued when computing a timestamp proposal $t$ (\refline{algo:commit:fun-ts}).
This function generates a single attached promise for the proposal $t$, stored in a mapping $\Attached$ (\refline{algo:commit:attached}).
The function also generates detached promises for the timestamps ranging from $\Clock + 1$ up to $t - 1$ (\refline{algo:commit:detached}):
since the process bumps the $\Clock$ to $t$ (\refline{algo:commit:fun-ts-bump}), it will never assign a timestamp in this range.
Detached promises are accumulated in the $\Detached$ set.
In \reftab{tab:fast_paths} $d)$, process \B generates an attached promise $\tup{\B, \clockval{6}}$, while \C generates $\tup{\C, \clockval{6}}$.
Process \B does not issue detached promises, since its $\Clock$ is bumped only by $1$, from $5$ to $\clockval{6}$.
However, process \C bumps its $\Clock$ by $5$, from $1$ to $\clockval{6}$, generating four detached promises: $\tup{\C, 2}$, $\tup{\C, 3}$, $\tup{\C, 4}$, $\tup{\C, 5}$.

\begin{algorithm}[t]
    \setcounter{AlgoLine}{\value{CommitLastLine}}
    \small

    \SubAlgo{\Periodically}{
        \Send $\MPromises(\Detached, \Attached)$ \To $\procs_p$ \label{algo:execution:mvotes-send}
    }

    \SubAlgo{\Receive $\MPromises(D, A)$\label{algo:execution:mvotes-recv}}{
        $C \leftarrow \bigunion \{ a \mid \tup{\id, a} \in A \land \id \in \commit \union \execute \}$ \label{algo:execution:learn-attached} \;
        $\Promises \leftarrow \Promises \union D \union C$ \label{algo:execution:learn-allowed-votes}
    }

    \SubAlgo{\Periodically\label{algo:execution:execute-handler}}{
        \mbox{$h \leftarrow \sort \{ \FunHCV(j) \mid j \in \procs_p \}$} \label{algo:execution:sort} \;
        $\mathit{ids} \leftarrow \{ \id \in \commit \mid \ts[\id] \leq h[\hspace{1pt}\floor{\frac{r}{2}}\hspace{1pt}] \}$ \label{algo:execution:stable-and-committed} \;
        \For{$\id \in \mathit{ids}$ \OrderedBy $\tup{\ts[\id], \id}$ \label{algo:execution:order}}{
            $\exec_p(\cmd[\id])$;
            $\phase[\id] \leftarrow \EXECUTE$ \label{algo:execution:execute-phase}
        }
    }

    \SubAlgo{$\FunHCV (j)$}{
        $\max\{c \in \nat \mid \forall u \in \{1 \dots c \} \cdot \tup{j, u} \in \Promises \}$ \label{algo:execution:votes}
    }
    \newcounter{ExecutionLastLine}
    \setcounter{ExecutionLastLine}{\value{AlgoLine}}

    \caption{
        Execution protocol at process $i \in \procs_p$.
    }
    \label{algo:execution}
\end{algorithm}

\refalgo{algo:execution} specifies the \SYS execution protocol at a process replicating a partition $p$.
Periodically, each process broadcasts its detached and attached promises to the
other processes replicating the same partition by sending them in an $\MPromises$ message (\refline{algo:execution:mvotes-send})\footnote{To minimize the size of these messages, a promise is sent only once in the absence of failures. Promises can be garbage-collected as soon as they are received by all the processes within the partition.}.
When a process receives the promises (\refline{algo:execution:mvotes-recv}), it adds them to a set $\Promises$.
Detached promises are added immediately.
An attached promise associated with a command identifier $\id$ is only added once
$\id$ is committed or executed (\refline{algo:execution:learn-attached}).

\subsubsection*{Stability detection}
\SYS determines when a timestamp is stable (\refprop{prop:stability}) according to the following theorem.
\begin{theoremsc}
  \label{theorem:stable}
  \upshape
  A timestamp $s$ is stable at a process $i$ if the variable $\Promises$ contains all the promises up to $s$ by some set of processes $Q$ with $\setsize{Q} \geq \floor{\frac{r}{2}} + 1$.
\end{theoremsc}
\begin{proof}
  Assume that at some time $\tau$ the variable $\Promises$ at a process $i$ contains all the promises up to $s$ by some set of processes $Q$ with $\setsize{Q} \geq \floor{\frac{r}{2}} + 1$.
  Assume further that a command $c$ with identifier $\id$ is eventually committed with timestamp $t \leq s$ at some process $j$, i.e., $j$ receives an $\MCommit(\id, t)$.
  We need to show that command $c$ is committed at $i$ at time $\tau$.
  By \refprop{prop:bump-at-majority} we have $t = \max\{ t_k \mid k \in Q' \}$, where $\setsize{Q'} \geq \floor{\frac{r}{2}} + 1$ and $t_k$ is the output of function $\FunTs(\id, \_)$ at a process $k$.
  As $Q$ and $Q'$ are majorities, there exists some process $l \in Q \cap Q'$.
  Then this process attaches a promise $\tup{l, t_l}$ to $c$ (\refline{algo:commit:attached}) and $t_l \leq t \leq s$.
  Since the variable $\Promises$ at process $i$ contains all the promises up to $s$ by process $l$, it also contains the promise $\tup{l, t_l}$.
  According to \refline{algo:execution:learn-attached}, when this promise is incorporated into $\Promises$, command $c$ has been already committed at $i$, as required.
\end{proof}

A process periodically computes the highest contiguous promise for each process replicating the same partition, and stores these promises in a sorted array $h$ (\refline{algo:execution:sort}).
It determines the highest stable timestamp according to \reftheorem{theorem:stable} as the one at index $\floor{\frac{r}{2}}$ in $h$.
The process then selects all the committed commands with a timestamp no higher than the stable one and executes them in the timestamp order, breaking ties using their identifiers.
After a command is executed, it is moved to the $\EXECUTE$ phase, which ends its journey. %

\begin{figure}[t]
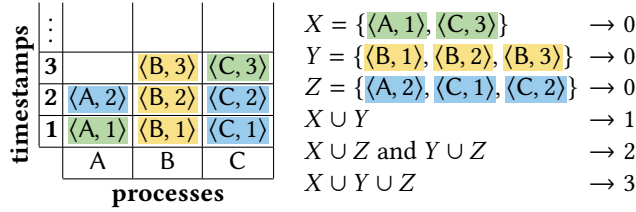

    \begin{minipage}{.37\linewidth}
        \begin{center}
        \setlength\tabcolsep{.19em}
        \begin{tabular}{cccccc}
            \tslabel  & \tbfentry{$\vdots$} & \tentry{}      & \tentry{}        & \tentry{}
            \\ \cline{2-5}
            \tentry{} & \tbfentry{3}        & \tentry{}      & \tentry{\bthree} & \tentry{\cthree}
            \\ \cline{2-5}
            \tentry{} & \tbfentry{2}        & \tentry{\atwo} & \tentry{\btwo}   & \tentry{\ctwo}
            \\ \cline{2-5}
            \tentry{} & \tbfentry{1}        & \tentry{\aone} & \tentry{\bone}   & \tentry{\cone}
            \\ \cline{2-5}
                      & \tbfentry{}         & \tentry{\A}    & \tentry{\B}      & \tentry{\C}
            \\ \cline{3-5}
                      &                     & \procslabel
        \end{tabular}
        \end{center}
    \end{minipage}%
    \begin{minipage}{.06\linewidth}
        \hfill
    \end{minipage}%
    \begin{minipage}{.6\linewidth}
        \begin{center}
        \setlength\tabcolsep{.19em}
        \begin{tabular}{lcccccclcc}
            $\votesgreen = \{ \aone, \cthree \}$                              & $ \rightarrow 0$
            \\
            $\votesred = \{ \bone, \btwo, \bthree \}$                         & $ \rightarrow 0$
            \\
            $\votesblue = \{ \atwo, \cone, \ctwo \}$                          & $ \rightarrow 0$
            \\
            $\votesgreen \union \votesred$                                    & $\rightarrow 1$
            \\
            $\votesgreen \union \votesblue$ and $\votesred \union \votesblue$ & $\rightarrow 2$
            \\
            $\votesgreen \union \votesred \union \votesblue $                 & $\rightarrow 3$
        \end{tabular}
        \end{center}
    \end{minipage}

    \vspace{-0.25cm}
    \caption{
        Stable timestamps for different sets of promises.
    }
    \label{fig:votes_table}
    \Description{TODO}
    \vspace{-0.30cm}
\end{figure}

To gain more intuition about the above mechanism, consider \reffig{fig:votes_table}, where $r = 3$.
There we represent the variable $\Promises$ of some process as a table, with processes as columns and timestamps as rows.
For example, a promise $\tup{\A, 2}$ is in $\Promises$ if it is present in column $\A$, row $2$.
There are three sets of promises, $\votesgreen$, $\votesred$ and $\votesblue$, to be added to $\Promises$.
For each combination of these sets, the right hand side of \reffig{fig:votes_table} shows the highest stable timestamp if all the promises in the combination are in $\Promises$.
For instance, assume that $\Promises = \votesred \cup \votesblue$, so that the set contains promise $2$ by \A, all promises up to $3$ by \B, and all promises up to $2$ by \C.
As $\Promises$ contains all promises up to $2$ by the majority $\{ \B, \C \}$, timestamp $2$ is stable: any uncommitted command $c$ must be committed with a timestamp higher than $2$.
Indeed, since $c$ is not yet committed, $\Promises$ does not contain any promise attached to $c$ (\refline{algo:execution:learn-attached}).
Moreover, to get committed, $c$ must generate attached promises at a majority of processes (\refprop{prop:bump-at-majority}), and thus, at either \B or \C.
If $c$ generates an attached promise at \B, its coordinator will receive at least proposal $4$ from \B; if at \C, its coordinator will receive at least proposal $3$.
In either case, and since the committed timestamp is the highest timestamp proposal, the committed timestamp of $c$ must be at least $3 > 2$, as required.

In our implementation, promises generated by fast-quorum processes
when computing their proposal for a command (\refline{algo:commit:fun-ts}) \pagebreak{}
are piggybacked on the $\MProposeAck$ message, and then broadcast by the
coordinator in the $\MCommit$ message (omitted from the pseudocode).
This speeds up stability detection and often allows a timestamp of a command to
become stable immediately after it is decided.
Notice that when committing a command, \SYS generates detached promises up to the timestamp of that command (\refline{algo:commit:mcommit-bump}).
This helps ensuring the liveness of the execution mechanism, since the propagation of these promises contributes to advancing the highest stable timestamp.

\subsection{Timestamp Stability vs Explicit Dependencies}
\label{sec:sys_comparison}

\newcommand\cmdw{\scolorbox{lightgreen}{w}}
\newcommand\cmdx{\scolorbox{lightsilver}{x}}
\newcommand\cmdy{\scolorbox{lightorange}{y}}
\newcommand\cmdz{\scolorbox{lightblue}{z}}

Prior leaderless protocols~\cite{epaxos,bpaxos,gryff,atlas,caesar} commit each command $c$ with a set of {\em explicit dependencies} $\dep[c]$.
In contrast, \SYS does not track explicit dependencies, but uses timestamp stability to decide when to execute a command.
This allows \SYS to ensure progress under synchrony.
Protocols using explicit dependencies do not offer such a guarantee, as they can arbitrarily delay the execution of a command.
In practice, this translates into a high tail latency.

\begin{figure}[t]
    \begin{minipage}{.5\linewidth}
        \begin{center}
            \setlength\tabcolsep{.19em}
            \begin{tabular}{ccccccclcc}
                \tslabel  & \tbfentry{$\vdots$} & \tentry{}            & \tentry{}          & \tentry{}          &
                \\ \cline{2-5}
                \tentry{} & \tbfentry{3}        & \tentry{\compathree} & \tentry{}          & \tentry{}          &
                \\ \cline{2-5}
                \tentry{} & \tbfentry{2}        & \tentry{}            & \tentry{\compbtwo} & \tentry{\compctwo} &
                \\ \cline{2-5}
                \tentry{} & \tbfentry{1}        & \tentry{\compaone}   & \tentry{\compbone} & \tentry{\compcone} &
                \\ \cline{2-5}
                          & \tbfentry{}         & \tentry{\A}          & \tentry{\B}        & \tentry{\C}        &
                \\ \cline{3-5}
                          &                     & \procslabel          &                    &                    &
            \end{tabular}
        \end{center}
    \end{minipage}%
    \begin{minipage}{.45\linewidth}
        \vspace{-0.35em}
        \begin{center}
            \begin{tabular}{c}
                \begin{xy}
                    <0.5cm,0pt>:
                    (-3,2)*+{\scolorbox{lightgreen}{w}}="w";
                    (-1,2)*+{\scolorbox{lightorange}{y}}="y";
                    (1,2)*+{\scolorbox{lightblue}{z}}="z";
                    (3,2)*+{\scolorbox{lightsilver}{x}}="x";
                    (-2.9,1.0)*+{}="legend-b";
                    (-1.4,1.0)*+{}="legend-a";
                    (0.6,1.0)*+{\text{``depends on''}}="";
                    \ar@{->} "y"; "w";
                    \ar@{->} "z"; "y";
                    \ar@(ul,ur) "w"; "z";
                    \ar@{->} "x"; "z";
                    \ar@{->} "legend-a"; "legend-b";
                \end{xy}
            \end{tabular}
        \end{center}
        \vspace{-0.4em}
        \hrule height 0.04cm
        \vspace{0.1em}
        \begin{center}
            \begin{tabular}{c}
                \begin{xy}
                    <0.5cm,0pt>:
                    (-3,2)*+{\scolorbox{lightgreen}{w}}="w";
                    (-1,2)*+{\scolorbox{lightorange}{y}}="y";
                    (1,2)*+{\scolorbox{lightblue}{z}}="z";
                    (3,2)*+{\scolorbox{lightsilver}{x}}="x";
                    (-2.9,1.0)*+{}="legend-b";
                    (-1.4,1.0)*+{}="legend-a";
                    (0.6,1.0)*+{\text{``blocked on''}}="";
                    \ar@{.>} "y"; "w";
                    \ar@{.>} "z"; "y";
                    \ar@{.>} "x"; "z";
                    \ar@{.>} "legend-a"; "legend-b";
                \end{xy}
            \end{tabular}
        \end{center}
    \end{minipage}
    \vspace{-0.4cm}
    \caption{
        Comparison between timestamp stability (left) and two approaches using explicit dependencies (right).
    }
    \label{fig:comparison}
    \Description{TODO}
    \vspace{-0.30cm}
\end{figure}
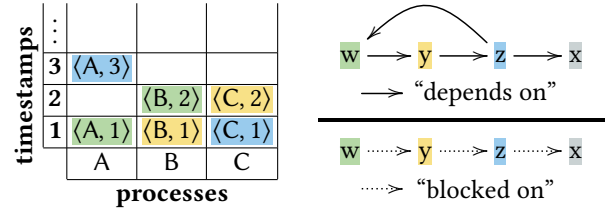

\reffig{fig:comparison} illustrates this issue using four commands $w, x, y, z$ and $r = 3$ processes.
Process \A submits $\cmdw$ and $\cmdx$, \B submits $\cmdy$, and \C submits $\cmdz$.
Commands arrive at the processes in the following order:
$\cmdw$, $\cmdx$, $\cmdz$ at \A;
$\cmdy$, $\cmdw$ at \B;
and $\cmdz$, $\cmdy$ at \C.
Because in this example only process \A has seen command $x$, this command is not yet committed.
In \SYS, the above command arrival order generates the following attached promises:
$\{ \compaone, \compbtwo \}$ for $w$,
$\{ \compatwo \}$ for $x$,
$\{ \compbone, \compctwo \}$ for $y$, and
$\{ \compcone, \compathree \}$ for $z$.
Commands $w$, $y$ and $z$ are then committed with the following timestamps:
$\ts[w] = 2$,
$\ts[y] = 2$, and
$\ts[z] = 3$.
On the left of \reffig{fig:comparison} we present the $\Promises$ variable of some process once it receives the promises attached to the three committed commands.
Given these promises, timestamp $2$ is stable at the process.
Even though command $x$ is not committed, timestamp stability ensures that its timestamp must be greater than $2$.
Thus, commands $w$ and $y$, committed with timestamp $2$, can be safely executed.
We now show how two approaches that use explicit dependencies behave in the above example.

\subsubsection*{Dependency-based ordering}
EPaxos~\cite{epaxos} and follow-ups~\cite{bpaxos,gryff,atlas} order commands based on their committed dependencies.
For example, in EPaxos, the above command arrival order results in commands $w$, $y$ and $z$ committed with the following dependencies:
$\dep[w] = \{ y \}$,
$\dep[y] = \{ z \}$,
$\dep[z] = \{ w, x \}$.
These form the graph shown on the top right of \reffig{fig:comparison}.
Since the dependency graph may be cyclic (as in \reffig{fig:comparison}), commands cannot be simply executed in the order dictated by the graph.
Instead, the protocol waits until it forms strongly connected components of the graph and then executes these components one at a time.
As we show in~\tra{\ref{sec:app_issues}}{D}, the size of such components is a priori unbounded.
This can lead to pathological scenarios where the protocol continuously commits commands but can never execute them, even under a synchronous network~\cite{leaderless,epaxos}.
It may also significantly delay the execution of committed commands, as illustrated by our example: since command $x$ has not yet been committed, and the strongly connected component formed by the committed commands $w$, $y$ and $z$ depends on $x$, no command can be executed -- unlike in \SYS.
As we demonstrate in our experiments (\refsec{sec:evaluation}), execution delays in such situations lead to high tail latencies.

\subsubsection*{Dependency-based stability}
Caesar~\cite{caesar} associates each command $c$ not only with a set of dependencies $\dep[c]$, but also with a unique timestamp $\ts[c]$.
Commands are executed in timestamp order, and dependencies are used to determine when a timestamp is stable, and thus when the command can be executed.
For this, dependencies have to be consistent with timestamps in the following sense: for any two commands $c$ and $c'$, if $\ts[c] < \ts[c']$, then $c \in \dep[c']$.
Then the timestamp of a command can be considered stable when the transitive dependencies of the command are committed.

Caesar determines the predecessors of a command while agreeing on its timestamp.
To this end, the coordinator of a command sends the command to a quorum together with a timestamp proposal.
The proposal is committed when enough processes vote for it.
Assume that in our example \A proposes $w$ and $x$ with timestamps $1$ and $4$, respectively, \B proposes $y$ with $2$, and \C proposes $z$ with $3$.
When \B receives command $w$ with timestamp proposal $1$, it has already proposed $y$ with timestamp $2$.
If these proposals succeed and are committed, the above invariant is maintained only if $w$ is a dependency of $y$.
However, because $y$ has not yet been committed, its dependencies are unknown and thus \B cannot yet ensure that $w$ is a dependency of $y$.
For this reason, \B must block its response about $w$ until $y$ is committed.
Similarly, command $y$ is blocked at \C waiting for $z$, and $z$ is blocked at \A waiting for $x$.
This situation, depicted in the bottom right of \reffig{fig:comparison}, results in no command being committed -- again, unlike in \SYS.
In fact, as we show in~\tra{\ref{sec:app_issues}}{D}, the blocking mechanism of Caesar allows pathological scenarios where commands are never committed at all.
Similarly to EPaxos, in practice this leads to high tail latencies (\refsec{sec:evaluation}).
In contrast to Caesar, \SYS computes the predecessors of a command separately from agreeing on its timestamp, via background stability detection.
This obviates the need for artificial delays in agreement, allowing \SYS to offer low tail latency (\refsec{sec:evaluation}).

\subsubsection*{Limitations of timestamp stability}
Protocols that track explicit dependencies are able to distinguish between read and
write commands.
In these protocols writes depend on both reads and writes, but reads only
have to depend on writes.
The latter feature improves the performance in read-dominated workloads.
In contrast, \SYS does not distinguish between read and write
commands, so that its performance is not affected by the ratio of reads
in the workload.
We show in \refsec{sec:evaluation} that this limitation does not prevent \SYS
from providing similar throughput as the best-case scenario (i.e., a read-only workload)
of protocols such as EPaxos and Janus.
Adapting techniques that exploit the distinction between reads and writes
is left as future work.

\section{Multi-Partition Protocol}
\label{sec:sys_partial}

\refalgo{algo:partial} extends the \SYS commit and execution protocols to handle commands that access multiple partitions.
This is achieved by submitting a multi-partition command at each of the
partitions it accesses using \refalgo{algo:commit}.
Once committed with some timestamp at each of these partitions, the command's
final timestamp is computed as the maximum of the committed timestamps.
A command is executed once it is stable at all the partitions it accesses.
As previously, commands are executed in the timestamp order.

In more detail, when a process $i$ submits a multi-partition command $c$ on behalf of a client (\refline{algo:commit:submit}), it sends an $\MSubmit$ message to a set $\procs_c^i$.
For each partition $p$ accessed by $c$, the set $\procs_c^i$ contains a responsive replica of $p$ close to $i$ (e.g., located in the same data center).
The processes in $\procs_c^i$ then serve as coordinators of $c$ in the respective partitions, following the steps in \refalgo{algo:commit}.
This algorithm ends with the coordinator in each partition sending an $\MCommit$ message to $\procs_c$, i.e., all processes that replicate a partition accessed by $c$ (lines~\ref{algo:commit:fast-path} and~\ref{algo:commit:consensus-end}; note that $\procs_c \neq \procs_p$ because $c$ accesses multiple partitions).
Hence, each process in $\procs_c$ receives as many $\MCommit$s as the number of partitions accessed by $c$.
Once this happens, the process executes the handler at line~\ref{algo:partial:mcommit-recv} in \refalgo{algo:partial}, which replaces the previous $\MCommit$ handler in \refalgo{algo:commit}.
The process computes the final timestamp of the multi-partition command as the highest of the timestamps
committed at each partition, moves the command to the $\COMMIT$ phase and bumps the $\Clock$
to the computed timestamp, generating detached promises.

Commands are executed using the handler at line~\ref{algo:partial:execute-handler}, which replaces that at line~\ref{algo:execution:execute-handler}.
This detects command stability using \reftheorem{theorem:stable}, which also holds in the multi-partition case.
The handler signals that a command $c$ is stable at a partition by sending an $\MStable$ message (line~\ref{algo:partial:stable-send}).
Once such a message is received from all the partitions accessed by $c$, the command is executed.
The exchange of $\MStable$ messages follows the approach in~\cite{ssmr} and ensures and the real-time order constraint in the Ordering property of PSMR (\refsec{sec:model}).

\begin{algorithm}[t]
    \setcounter{AlgoLine}{\value{ExecutionLastLine}}
    \small

    \SubAlgo{\Receive $\MCommit(\id, t_j)$ \From $j \in \procs_{\cmd[\id]}^i$\label{algo:partial:mcommit-recv}}{
        \Pre $\id \in \pending$ \;
        $\ts[\id] \leftarrow \max \{ t_j \mid j \in P \}$; \label{algo:partial:final-ts}
        $\phase[\id] \leftarrow \COMMIT$ \label{algo:partial:commit-phase} \;
        $\FunBump(\ts[\id])$ \label{algo:partial:mcommit-bump}
    }

    \SubAlgo{\Periodically\label{algo:partial:execute-handler}}{
        \mbox{$h \leftarrow \sort \{ \FunHCV(j) \mid j \in \procs_p \}$} \label{algo:partial:sort} \;
        $\mathit{ids} \leftarrow \{ \id \in \commit \mid \ts[\id] \leq h[\hspace{1pt}\floor{\frac{r}{2}}\hspace{1pt}] \}$ \label{algo:partial:stable-and-committed} \;
        \For{$\id \in \mathit{ids}$ \OrderedBy $\tup{\ts[\id], \id}$ \label{algo:partial:order}}{
            \Send $\MStable(\id)$ \To $\procs_{\cmd[\id]}$ \label{algo:partial:stable-send} \;
            \mbox{\textbf{wait} \Receive $\MStable(\id)$ \From $\forall j \in \procs_{\cmd[\id]}^i$} \label{algo:partial:stable-recv} \;
            $\exec_p(\cmd[\id])$;
            $\phase[\id] \leftarrow \EXECUTE$ \label{algo:partial:execute-phase}
        }
    }

    \SubAlgo{\Receive $\MPropose(\id, c, \qs, t)$ \From $j$ \label{algo:partial:mpropose-recv}}{
        \nonl $\dots$ \;
        \Send $\MBump(\id, \ts[\id])$ \To $\procs_c^i$ \label{algo:partial:mbump-send}
    }

    \SubAlgo{\Receive $\MBump(\id, t)$ \label{algo:partial:mbump-recv}}{
        \Pre $\id \in \propose$ \;
        $\FunBump(t)$ \label{algo:partial:mbump-bump}
    }

    \newcounter{PartialLastLine}
    \setcounter{PartialLastLine}{\value{AlgoLine}}

    \caption{
        Multi-partition protocol at process $i \in \procs_p$.
    }
    \label{algo:partial}
\end{algorithm}

\subsubsection*{Example}
\reffig{fig:partial-deployment} shows an example of \SYS $f = 1$ with $r = 5$ and $2$ partitions.
Only 3 processes per partition are depicted.
Partition $0$ is replicated at \A, \B and \C, and partition $1$ at \F, \G and \H.
Processes with the same color (e.g., \B and \G) are located nearby each other (e.g., in the same machine or data center).
Process \A and \F are the coordinators for some command that accesses the two partitions.
At partition $0$, \A computes $\clockval{6}$ as its timestamp proposal and sends
it in an $\MPropose$ message to the fast quorum $\{ \A, \B, \C \}$
(the downward arrows in \reffig{fig:partial-deployment}).
These processes make the same proposal, and thus the command is committed
at partition $0$ with timestamp $6$.
Similarly, at partition $1$, \F computes $\clockval{10}$ as its proposal and sends it to
$\{ \F, \G, \H \}$, all of which propose the same.
The command is thus committed at partition $1$ with timestamp $10$.
The final timestamp of the command is then computed as $\max\{6, 10\} = 10$.

Assume that the stable timestamp at \A is $5$ and at \F is $9$ when they compute
the final timestamp for the command.
Once \F receives the attached promises by the majority $\{ \F, \G, \H \}$,
timestamp $10$ becomes stable at \F.
This is not the case at \A, as the attached promises by the majority
$\{ \A, \B, \C \}$ only make timestamp $6$ stable.
However, processes \A, \B and \C also generate detached promises up to timestamp
$10$ when receiving the $\MCommit$ messages for the command
(\refline{algo:partial:mcommit-bump}).
When \A receives these promises, it declares timestamp $10$ stable.
This occurs after two extra message delays: an $\MCommit$ from \A
and \F to \B and \C, and then $\MPromises$ from \B and \C back to \A.
Since the command's timestamp is stable at both \A and \F, once these processes exchange $\MStable$ messages, the command can finally be executed at each.

\subsubsection*{Faster stability}
\SYS avoids the above extra delays by generating the detached promises
needed for stability earlier than in the $\MCommit$ handler. For this
we amend the $\MPropose$ handler as shown
in \refalgo{algo:partial}. When a process receives an $\MPropose$ message, it
follows the same steps as in \refalgo{algo:commit}. It then
additionally sends an $\MBump$ message containing its proposal to the
nearby processes that replicate a partition accessed by the command
(\refline{algo:partial:mbump-send}). Upon receiving this message
(\refline{algo:partial:mbump-recv}), a process bumps its $\Clock$ to the
timestamp in the message, generating detached promises.

\tikzstyle{regiona} = [circle, draw=orange, fill=lightorange, very thick, minimum size=7mm]
\tikzstyle{regionb} = [circle, draw=green, fill=lightgreen, very thick, minimum size=7mm]
\tikzstyle{regionc} = [circle, draw=blue, fill=lightblue, very thick, minimum size=7mm]
\tikzstyle{detached} = [dashed, draw=silver, fill=silver!10, very thick]

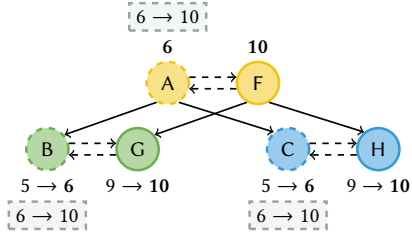
\begin{figure}[t]
    \centering
    \scalebox{.8}{
        \begin{tikzpicture}
            \node[regiona,dashed] (a) at (0, 0) {\A};
            \node at (0, 0.6) {\clockval{6}};
            \node[detached] at (0, 1.1) {$6 \rightarrow 10$};
            \node[regiona] (f) at (1.5, 0) {\F};
            \node at (1.5, 0.6) {\clockval{10}};

            \node[regionb,dashed] (b) at (-2, -1.1) {\B};
            \node at (-2, -1.7) {$5 \rightarrow \clockval{6}$};
            \node[detached] at (-2, -2.2) {$6 \rightarrow 10$};
            \node[regionb] (g) at (-0.5, -1.1) {\G};
            \node at (-0.5, -1.7) {$9 \rightarrow \clockval{10}$};

            \node[regionc,dashed] (c) at (2, -1.1) {\C};
            \node at (2, -1.7) {$5 \rightarrow \clockval{6}$};
            \node[detached] at (2, -2.2) {$6 \rightarrow 10$};
            \node[regionc] (h) at (3.5, -1.1) {\H};
            \node at (3.5, -1.7) {$9 \rightarrow \clockval{10}$};

            \draw[->,thick] (a.240) -- (b.40);
            \draw[->,thick] (a.300) -- (c.130);
            \draw[->,thick] (f.240) -- (g.40);
            \draw[->,thick] (f.300) -- (h.130);

            \draw[->,thick,dashed] (a.15) -- (f.165);
            \draw[->,thick,dashed] (f.195) -- (a.-15);

            \draw[->,thick,dashed] (b.15) -- (g.165);
            \draw[->,thick,dashed] (g.195) -- (b.-15);

            \draw[->,thick,dashed] (c.15) -- (h.165);
            \draw[->,thick,dashed] (h.195) -- (c.-15);
        \end{tikzpicture}
    }
    \vspace{-0.2cm}
    \caption{
      Example of \SYS with 2 partitions.
      Next to each process we show the clock updates upon receiving $\MPropose$
      messages and, in dashed boxes, the updates upon receiving $\MCommit$ or $\MBump$ messages (whichever
      occurs first).  }
    \label{fig:partial-deployment}
    \Description{TODO}
    \vspace{-0.30cm}
\end{figure}

In \reffig{fig:partial-deployment}, $\MBump$ messages are depicted by horizontal
dashed arrows. When \G computes its proposal $10$, it sends an $\MBump$ message
containing $10$ to process \B.  Upon reception, \B bumps its $\Clock$ to $10$,
generating detached promises up to that value. The same happens at \A and
\C.  Once the detached promises by the majority $\{ \A, \B, \C \}$ are known at
\A, the process again declares $10$ stable. In this
case, \A receives the required detached promises in two message
delays earlier than when these promises are generated via $\MCommit$. This strategy often reduces the number of
message delays necessary to execute a multi-partition command. However, it is not always sufficient (e.g., imagine that \H proposed $11$ instead of $10$), and
thus, the promises issued in the $\MCommit$ handler (\refline{algo:partial:mcommit-bump}) are still necessary for multi-partition commands.

\subsubsection*{Genuineness and parallelism}
The above protocol is {\em genuine}: for every command $c$, only the processes in $\procs_c$ take steps to order and execute $c$~\cite{Guerraoui2001}.
This is not the case for existing leaderless protocols for partial replication, such as Janus~\cite{janus}.
With a genuine protocol, partitioning the application state brings scalability in parallel workloads: an increase in the number of partitions (and thereby of available machines) leads to an increase in throughput.
When partitions are colocated in the same machine, the message passing in \refalgo{algo:partial} can be optimized and replaced by shared-memory operations.
Since \SYS runs an independent instance of the protocol for each partition replicated at the process,
the resulting protocol is highly parallel.

\section{Recovery Protocol}
\label{sec:sys_recovery}

The initial coordinator of a command at some partition $p$ may fail or be slow to respond, in which
case \SYS allows a process to take over its role and recover the command's
timestamp.
We now describe the protocol \SYS follows in this case, which is inspired by that of Atlas~\cite{atlas}.
This protocol at a process $i \in \procs_p$ is given in \refalgo{algo:recovery}.
We use $\initial_p(\id)$ to denote a function that extracts from the command identifier $\id$ its initial coordinator at partition $p$.

A process takes over as the coordinator for some command with identifier $\id$
by calling function $\FunRecover(\id)$ at \refline{algo:recovery:fun-recover}.
Only a process with $\id \in \pending$ can take over as a coordinator (\refline{algo:recovery:fun-recover-pre}): this ensures that the process knows the command payload and fast quorums.
In order to find out if a decision on the timestamp of $\id$ has been reached in
consensus, the new coordinator first performs an analog of Paxos Phase 1.
It picks a ballot number it owns higher than any it participated in so far
(\refline{algo:recovery:my-ballot}) and sends an $\MRec$ message with this
ballot to all processes.

As is standard in Paxos, a process accepts an $\MRec$ message only if the ballot in the message is greater than its $\bal[\id]$ (\refline{algo:recovery:mrec-pre}).
If $\bal[\id]$ is still $0$ (\refline{algo:recovery:mrec-first-see}), the process checks the command's phase to decide if it should compute its timestamp proposal for the command.
If $\phase[\id] = \PAYLOAD$ (\refline{algo:recovery:mrec-at-payload}), the process has not yet computed a timestamp proposal, and thus it does so at \refline{algo:recovery:mrec-proposal}.
It also sets the command's phase to $\RECOVERR$, which records that the timestamp proposal was computed in the $\MRec$ handler.
Otherwise, if $\phase[\id] = \PROPOSE$ (\refline{algo:recovery:mrec-at-propose}),
the process has already computed a timestamp proposal at \refline{algo:commit:mpropose-proposal}.
In this case, the process simply sets the command's phase to $\RECOVERP$, which records that the timestamp proposal was computed in the $\MPropose$ handler.
Finally, the process sets $\bal[\id]$ to the new ballot
and replies with an $\MRecAck$ message containing the timestamp ($\ts$), the command's phase ($\phase$) and the ballot at which the timestamp was previously accepted in consensus ($\abal$).
Note that $\abal[\id] = 0$ if the process has not yet accepted any consensus proposal.
Also note that \reflines{algo:recovery:mrec-at-payload}{algo:recovery:mrec-at-propose} are exhaustive: these are the only possible phases when $\id \in \pending$ (\refline{algo:recovery:mrec-pre}) and $\bal[\id] = 0$ (\refline{algo:recovery:mrec-first-see}), as recovery phases have non-zero ballots (\refline{algo:recovery:mrec-new-bal}).

\begin{algorithm}[t]
    \setcounter{AlgoLine}{\value{PartialLastLine}}
    \small

    \SubAlgo{$\FunRecover(\id)$ \label{algo:recovery:fun-recover}}{
        \Pre $\id \in \pending$  \label{algo:recovery:fun-recover-pre} \;
        $b \leftarrow i + r (\floor{\frac{\bal[\id] - 1}{r}} + 1)$ \label{algo:recovery:my-ballot} \;
        \Send $\MRec(\id, b)$ \To $\procs_p$ \label{algo:recovery:mrec-send} \;
    }

    \SubAlgo{\mbox{\Receive $\MRec(\id, b)$ \From $j$\label{algo:recovery:mrec-recv}}}{
        \Pre $\id \in \pending \land \bal[\id] < b$ \label{algo:recovery:mrec-pre} \;

        \If{$\bal[\id] = 0$ \label{algo:recovery:mrec-first-see}}{
            \If{$\phase[\id] = \PAYLOAD$ \label{algo:recovery:mrec-at-payload}}{
                $\ts[\id] \leftarrow \FunTs(\id, 0)$ \label{algo:recovery:mrec-proposal} \;
                $\phase[\id] \leftarrow \RECOVERR$ \label{algo:recovery:recoverr-phase}
            }
            \ElseIf{$\phase[\id] = \PROPOSE$ \label{algo:recovery:mrec-at-propose}}{
                $\phase[\id] \leftarrow \RECOVERP$ \label{algo:recovery:recoverp-phase}
            }
        }
        $\bal[\id] \leftarrow b$ \label{algo:recovery:mrec-new-bal} \;
        \mbox{\Send $\MRecAck(\id, \ts[\id], \phase[\id], \abal[\id], b)$ \To $j$}
    }

    \SubAlgo{\Receive $\MRecAck(\id, t_j, \ph_j, \ab_j, b)$ \From $\forall j \in Q$\label{algo:recovery:mrec-ack-recv}}{
        \Pre $\bal[\id] = b \land \setsize{Q} = r - f$ \label{algo:recovery:mrec-ack-pre} \;
        \If{$\exists k \in Q \cdot \ab_k \not = 0$ \label{algo:recovery:previously-accepted}}{
            \Let $k$ be such that $\ab_k$ is maximal \label{algo:recovery:highest-ballot} \;
            \Send $\MConsensus(\id, t_k, b)$ \To $\procs_p$ \label{algo:recovery:highest-ballot-mconsensus}
        }
        \Else{
            $I \leftarrow Q \cap \quorums[\id][p]$ \label{algo:recovery:intersection} \;
            $s \leftarrow \initial_p(\id) \in I \lor \exists k \in I \cdot \ph_k = \RECOVERR$ \label{algo:recovery:coordinator-check} \;
            $Q' \leftarrow $ \leIf{$s$}{$Q$}{$I$} \label{algo:recovery:quorum-selection}
            $t \leftarrow \max \{ t_j \mid j \in Q' \}$ \label{algo:recovery:rec-max} \;
            \Send $\MConsensus(\id, t, b)$ \To $\procs_p$ \label{algo:recovery:first-rec-mconsensus}
        }
    }
    \newcounter{RecoveryLastLine}
    \setcounter{RecoveryLastLine}{\value{AlgoLine}}

    \caption{Recovery protocol at process $i \in \procs_p$.}
    \label{algo:recovery}
\end{algorithm}

In the $\MRecAck$ handler (\refline{algo:recovery:mrec-ack-recv}), the new
coordinator computes the command's timestamp given the information in the $\MRecAck$ messages
and sends it in an $\MConsensus$ message to all processes.
As in Flexible Paxos, the new coordinator waits for $r - f$ such messages.
This guarantees that, if a quorum of $f + 1$ processes accepted an $\MConsensus$
message with a timestamp (which could have thus been sent in an $\MCommit$
message), the new coordinator will find out about this timestamp.
To maintain \refprop{prop:agreement}, if any process previously accepted a
consensus proposal (\refline{algo:recovery:previously-accepted}), by the standard
Paxos rules~\cite{paxos,flexible-paxos}, the coordinator selects the proposal
accepted at the highest ballot (\refline{algo:recovery:highest-ballot}).

If no consensus proposal has been accepted before, the new coordinator first computes at \refline{algo:recovery:intersection} the set of processes $I$ that belong both to the recovery quorum $Q$ and the fast quorum $\quorums[\id][p]$.
Then, depending on whether the initial coordinator replied and in which handler the processes in $I$ have computed their timestamp proposal, there are two possible cases that we describe next.

\emph{1) The initial coordinator replies or some process in $I$ has computed its
timestamp proposal in the $\MRec$ handler ($s = {\sf true}$,
\refline{algo:recovery:coordinator-check}).}
In either of these two cases the initial coordinator could not have taken the fast.
If the initial coordinator replies ($\initial_p(\id) \in I$), then it
has not taken the fast path before receiving the $\MRec$ message from the new one,
as it would have $\id \in \commit \union \execute$ and the $\MRec$ precondition requires $\id \in \pending$ (\refline{algo:recovery:mrec-pre}).
It will also not take the fast path in the future, since when processing the
$\MRec$ message it sets the command's phase to $\RECOVERP$
(\refline{algo:recovery:recoverp-phase}), which invalidates the $\MProposeAck$
precondition (\refline{algo:commit:mpropose-ack-pre}).
On the other hand, even if the initial coordinator replies but some fast-quorum process in $I$ has computed its timestamp proposal in the $\MRec$ handler, the fast path will not be taken either.
This is because the command's phase at such a process is set to $\RECOVERR$
(\refline{algo:recovery:recoverr-phase}),
which invalidates the $\MPropose$ precondition (\refline{algo:commit:mpropose-pre}).
Then, since the $\MProposeAck$ precondition requires a reply from all fast-quorum processes, the initial coordinator will not take the fast path.
Thus, in either case, the initial coordinator never takes the fast path.
For this reason, the new coordinator can
choose the command's timestamp in any way, as long as it maintains
\refprop{prop:bump-at-majority}.
Since $|Q| = r - f \geq r - \floor{\frac{r-1}{2}} \geq \floor{\frac{r}{2}} + 1$,
the new coordinator has the output of $\FunTs$ by a majority of processes,
and thus it computes the command's timestamp with $\max$
(\refline{algo:recovery:rec-max}), respecting \refprop{prop:bump-at-majority}.

\emph{2) The initial coordinator does not reply and all processes in $I$ have computed their timestamp proposal in the $\MPropose$ handler
($s = {\sf false}$, \refline{algo:recovery:coordinator-check}).}
In this case the initial coordinator could have taken the fast path with some
timestamp $t = \max\{ t_j \mid j \in \quorums[\id][p] \}$ and, if it did, the new
coordinator must choose that same timestamp $t$.
Given that the recovery quorum $Q$ has size $r - f$ and the fast quorum
$\quorums[\id][p]$ has size $\floor{\frac{r}{2}} + f$, the set of processes
$I = Q \cap \quorums[\id][p]$ contains at
least $\floor{\frac{r}{2}}$ processes (distinct from the initial
coordinator, as it did not reply).
Furthermore, recall that the processes from $I$ have the command's phase set to $\RECOVERP$ (\refline{algo:recovery:recoverp-phase}), which invalidates the $\MPropose$
precondition (\refline{algo:commit:mpropose-pre}).
Hence, if the initial coordinator took the fast path, then each process
in $I$ must have processed its $\MPropose$ before the $\MRec$ of the new
coordinator, and reported in the latter the timestamp from the former.
Then using \refprop{prop:recovery}, the new coordinator recovers $t$ by selecting the highest timestamp
reported in $I$ (\refline{algo:recovery:rec-max}).

\subsubsection*{Additional liveness mechanisms.}
As is standard, to ensure the progress of recovery, \SYS nominates a single process to call $\FunRecover$  using a partition-wide failure detector~\cite{weakest-fd}, and ensures that this process picks a high enough ballot.
\SYS additionally includes a mechanism to ensure that, if a correct process receives an $\MPayload$ or an $\MCommit$ message, then all correct process do; this is also necessary for recovery to make progress.
For brevity, we defer a detailed description of these mechanisms to~\tra{\ref{sec:app_sys}}{B}.

\subsubsection*{Correctness}
We have rigorously proved that \SYS satisfies the PSMR specification (\refsec{sec:model}), even in case of failures. Due to space constraints, we defer the proof to~\tra{\ref{sec:app_correctness}}{C}.

\newcommand\plot[1]{
    \centering
    \includegraphics[width=\linewidth,keepaspectratio]{img/#1}
    \vspace{-0.70cm}
}

\section{Performance Evaluation}
\label{sec:evaluation}

In this section we experimentally evaluate \SYS in deployments with full replication (i.e., each partition is replicated at all processes) and partial replication.
We compare \SYS with Flexible Paxos (FPaxos)~\cite{flexible-paxos}, EPaxos~\cite{epaxos}, Atlas~\cite{atlas}, Caesar~\cite{caesar} and Janus~\cite{janus}.
FPaxos is a variant of Paxos that, like \SYS, allows selecting the allowed number of failures $f$ separately from the replication factor $r$: it uses quorums of size $f+1$ during normal operation and quorums of size $r-f$ during recovery.
EPaxos, Atlas and Caesar are leaderless protocols that track explicit dependencies (\refsec{sec:sys_comparison}).
EPaxos and Caesar use fast quorums of size $\floor{\frac{3r}{4}}$
and $\ceil{\frac{3r}{4}}$, respectively.
Atlas uses fast quorums of the same size as \SYS, i.e.,
$\floor{\frac{r}{2}} + f$.
Atlas also improves the condition EPaxos uses for taking the fast path: e.g., when $r = 5$ and $f = 1$, Atlas always processes commands via the fast path, unlike EPaxos.
To avoid clutter, we exclude the results for EPaxos from most of our plots since its performance is similar to (but never better than) Atlas $f = 1$.
Janus is a leaderless protocol that generalizes EPaxos to the setting of partial replication.
It is based on an unoptimized version of EPaxos whose fast quorums contain all replicas in a given partition.
Our implementation of Janus is instead based on Atlas, which yields quorums of the same size as \SYS and a more permissive fast-path condition.
We call this improved version Janus*.
This protocol is representative of the state-of-the-art for partial replication,
and the authors of Janus have already compared it extensively to prior approaches (including MDCC~\cite{mdcc}, Tapir~\cite{tapir} and 2PC over Paxos~\cite{spanner}).

\subsection{Implementation}
\label{sec:evaluation_implementation}

\definecolor{myblue}{RGB}{41, 128, 185}
\definecolor{myorange}{RGB}{230, 126, 34}

To improve the fairness of our comparison, all protocols are implemented in the same
framework which consists of 33K lines of Rust and contains common functionality necessary to implement and evaluate the protocols.
This includes a networking layer, an in-memory key-value store, \texttt{dstat} monitoring, and a set of benchmarks (e.g. YCSB~\cite{ycsb}).
The source code of the framework is available at \textbf{\color{myblue}{\href{https://github.com/vitorenesduarte/fantoch}{github.com/vitorenesduarte/fantoch}}}.

The framework provides three execution modes: cloud, cluster and simulator.
In the cloud mode, the protocols run in wide area on Amazon EC2.
In the cluster mode, the protocols run in a local-area network, with delays injected between the machines to emulate wide-area latencies.
Finally, the simulator runs on a single machine and computes the observed client latency in a given wide-area configuration when CPU and network bottlenecks are disregarded.
Thus, the output of the simulator represents the best-case latency for a given scenario.
Together with \texttt{dstat} measurements, the simulator allows us to determine if the latencies obtained in the cloud or cluster modes represent the best-case scenario for a given protocol or are the effect of some bottleneck.

\subsection{Experimental Setup}
\label{sec:evaluation_setup}

\subsubsection*{Testbeds}
As our first testbed we use Amazon EC2 with \textsf{c5.2xlarge} instances (machines with 8 virtual CPUs and 16GB of RAM).
Experiments span up to 5 EC2 regions, which we call \emph{sites}:
Ireland (\textsf{eu-west-1}),
Northern California (\textsf{us-west-1}),
Singapore (\textsf{ap-southeast-1}),
Canada (\textsf{ca-central-1}),
and São Paulo (\textsf{sa-east-1}).
The average ping latencies between these sites range from 72ms to 338ms; we defer precise numbers to~\traend{\ref{sec:app_evaluation}}{A}.
Our second testbed is a local cluster where we inject wide-area delays similar to those observed in EC2.
The cluster contains machines with 6 physical cores and 32GB of RAM connected by a 10GBit network.

\subsubsection*{Benchmarks}
We first evaluate full replication deployments (\refsec{sec:evaluation_full}) using a microbenchmark where each command carries a key of 8 bytes and (unless specified otherwise) a payload of 100 bytes.
Commands access the same partition when they carry the same key, in which case we say that they \emph{conflict}.
To measure performance under a conflict rate $\rho$ of commands, a client chooses key $0$ with a probability $\rho$, and some unique key otherwise.
We next evaluate partial replication deployments (\refsec{sec:evaluation_partial}) using YCSB+T~\cite{ycsbt}, a transactional version of the YCSB benchmark~\cite{ycsb}.
Clients are closed-loop and always deployed in separate machines located in the same regions as servers.
Machines are connected via 16 TCP sockets, each with a 16MB buffer. Sockets are flushed every 5ms or when the buffer is filled, whichever is earlier.

\subsection{Full Replication Deployment}
\label{sec:evaluation_full}

\subsubsection*{Fairness}
We first evaluate a key benefit of leaderless SMR, its fairness:
the fairer the protocol, the more uniformly it satisfies different sites.
We compare \SYS, Atlas and FPaxos when the protocols are deployed over 5 EC2 sites under two fault-tolerance levels: $f \in \{1, 2\}$.
We also compare with Caesar which tolerates $f = 2$ failures in this setting.
At each site we deploy 512 clients that issue commands with a low conflict rate (2\%).

\reffig{fig:fairness} depicts the per-site latency provided by each protocol.
The FPaxos leader site is Ireland, as we have determined that this site produces the fairest latencies.
However, even with this leader placement, FPaxos remains significantly unfair.
When $f = 1$, the latency observed by clients at the leader site is 82ms, while in São Paulo and Singapore it is 267ms and 264ms, respectively.
When $f = 2$, the clients in Ireland, São Paulo and Singapore observe respectively the latency of 142ms, 325ms and 323ms.
Overall, the performance at non-leader sites is up to \textsf{3.3x} worse than at the leader site.

Due to their leaderless nature, \SYS, Atlas and Caesar satisfy the clients much more uniformly.
With $f = 1$, \SYS and Atlas offer similar average latency -- 138ms for \SYS and 155ms for Atlas.
However, with $f = 2$ \SYS clearly outperforms Atlas -- 178ms versus 257ms.
Both protocols use fast quorums of size $\floor{\frac{r}{2}} + f$.
But because quorums for $f = 2$ are larger than for $f=1$, the size of the dependency sets in Atlas increases.
This in turn increases the size of the strongly connected components in execution (\refsec{sec:sys_comparison}).
Larger components result in higher average latencies, as reported in \reffig{fig:fairness}.
Caesar provides the average latency of 195ms, which is 17ms higher than \SYS $f = 2$.
Although Caesar and \SYS $f = 2$ have the same quorum size with $r = 5$, the blocking mechanism of Caesar delays commands in the critical path (\refsec{sec:sys_comparison}), resulting in slightly higher average latencies.
As we now demonstrate, both Caesar and Atlas have much higher tail latencies than \SYS.

\subsubsection*{Tail latency}
\begin{figure}[t]
    \plot{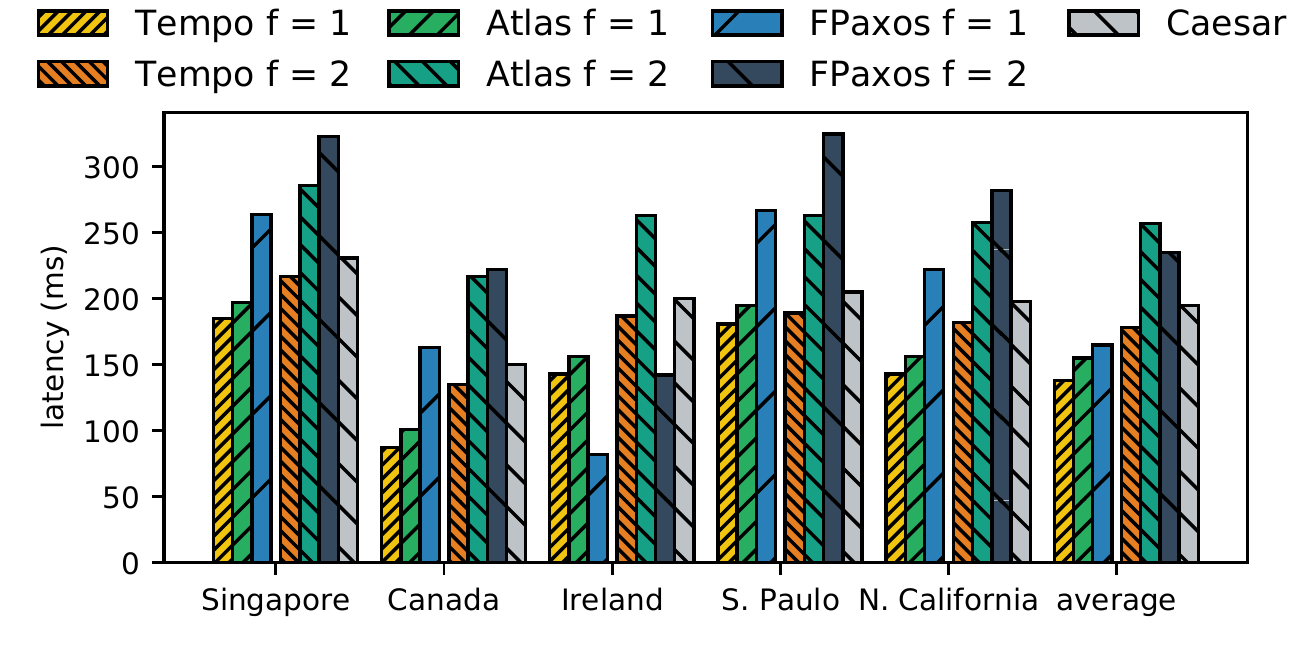}
    \caption{
        Per-site latency with 5 sites and 512 clients per site under a low
        conflict rate (2\%).
    }
    \label{fig:fairness}
    \Description{TODO}
    \vspace{-0.30cm}
\end{figure}

\begin{figure}[t]
    \plot{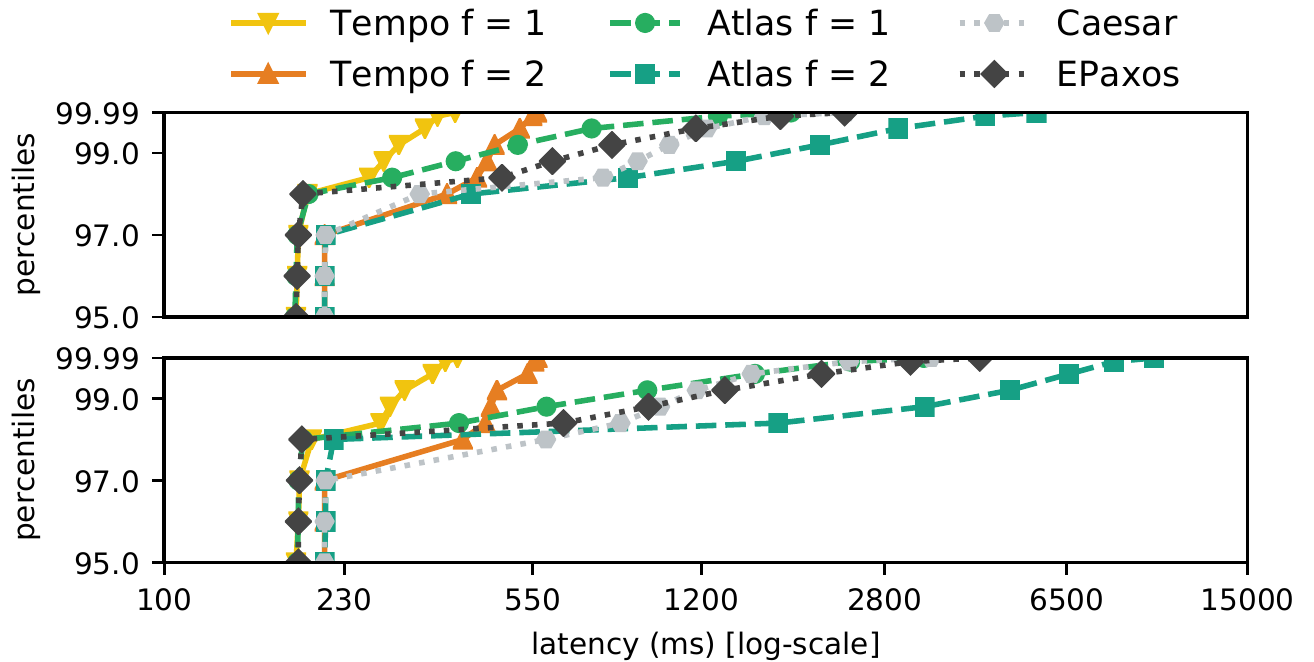}
    \caption{
        Latency percentiles with 5 sites and 256 (top) and 512 clients (bottom)
        per site under a low conflict rate (2\%).
    }
    \label{fig:tail_latency}
    \Description{TODO}
    \vspace{-0.30cm}
\end{figure}

\reffig{fig:tail_latency} shows the latency distribution of various protocols from the 95th to the 99.99th percentiles.
At the top we give results with 256 clients per site, and at the bottom with 512, i.e., the same load as in \reffig{fig:fairness}.

The tail of the latency distribution in Atlas, EPaxos and Caesar is very long.
It also sharply deteriorates when the load increases from 256 to 512 clients per site.
For Atlas $f = 1$, the 99th percentile increases from 385ms to 586ms while the 99.9th percentile increases from 1.3s to 2.4s.
The trend is similar for Atlas $f = 2$, making the 99.9th percentile increase from 4.5s to 8s.
The performance of EPaxos lies in between Atlas $f=1$ and Atlas $f=2$.
This is because with $5$ sites EPaxos has the same fast quorum size as Atlas $f = 1$, but takes the slow path with a similar frequency to Atlas $f = 2$.
For Caesar, increasing the number of clients also increases the 99th percentile from 893ms to 991ms and 99.9th percentile from 1.6s to 2.4s.
Overall, the tail latency of Atlas, EPaxos and Caesar reaches several seconds, making them impractical in these settings.
These high tail latencies are caused by ordering commands using explicit dependencies, which can arbitrarily delay command execution (\refsec{sec:sys_comparison}).

In contrast, \SYS provides low tail latency and predictable performance in both scenarios.
When $f = 1$, the 99th, 99.9th and 99.99th percentiles are respectively 280ms, 361ms and 386ms (averaged over the two scenarios).
When $f = 2$, these values are 449ms, 552ms and 562ms.
This represents an improvement of \textsf{1.4-8x} over Atlas, EPaxos and Caesar with 256 clients per site, and an improvement of \textsf{4.3-14x} with 512.
The tail of the distribution is much shorter with \SYS due to its efficient execution mechanism, which uses timestamp stability instead of explicit dependencies.

We have also run the above scenarios in our wide-area simulator.
In this case the latencies for Atlas, EPaxos and Caesar are up to 30\% lower, since CPU time is not accounted for.
The trend, however, is similar. This confirms that the latencies reported in \reffig{fig:tail_latency} accurately capture the effect of long dependency chains and are not due to a bottleneck in the execution mechanism of the protocols.

\subsubsection*{Increasing load and contention}

\begin{figure}[t]
    \vspace{-1.12em}
    \plot{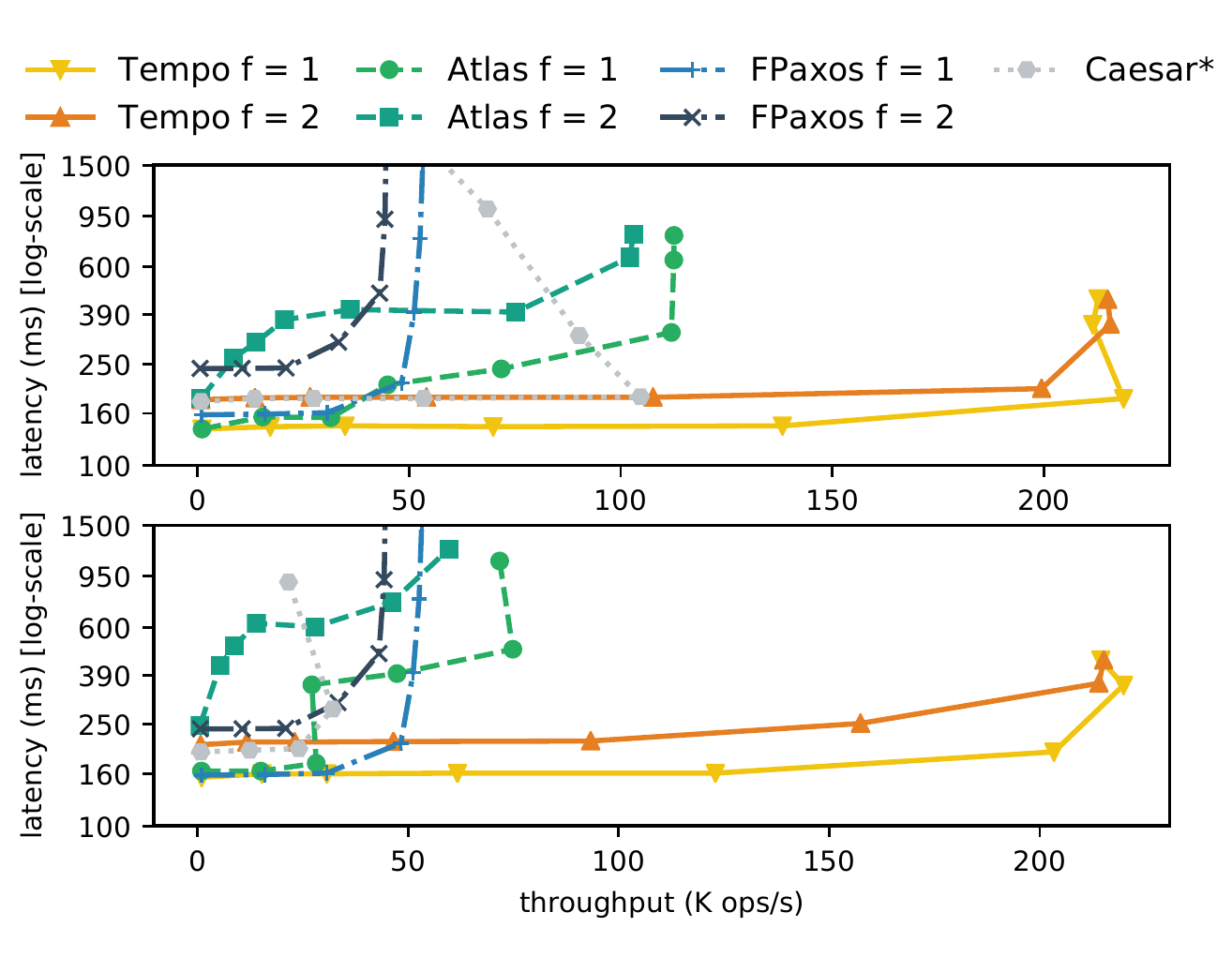}
    \vspace{-0.2em}
    \plot{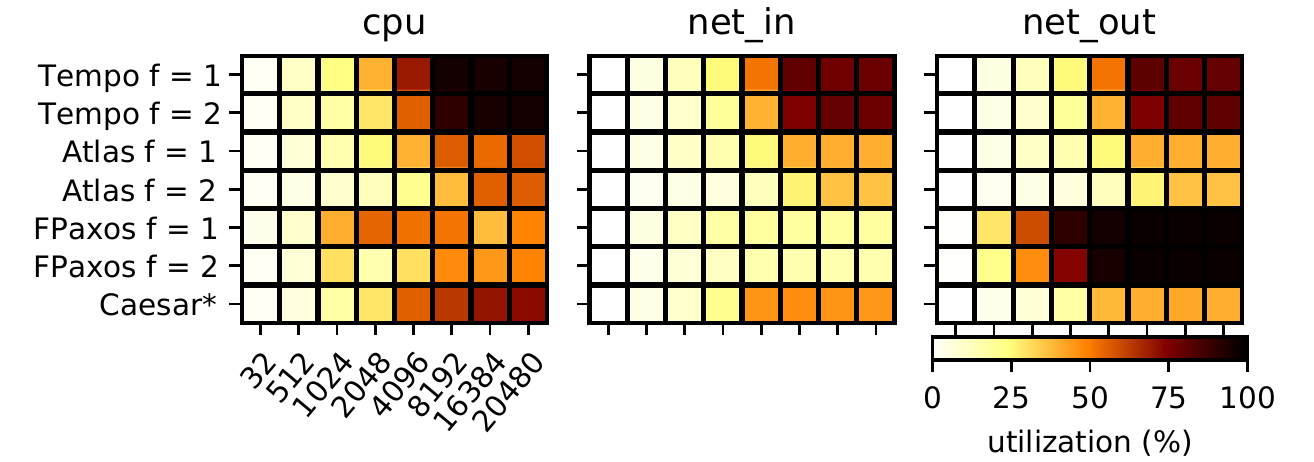}
    \vspace{-0.1em}
    \caption{
        Throughput and latency with 5 sites as the load increases from 32
        to 20480 clients per site under a low (2\% -- top) and moderate
        (10\% -- bottom) conflict rate.
        The heatmap shows the hardware utilization when the conflict rate is
        2\%.
    }
    \label{fig:increasing_load}
    \Description{TODO}
    \vspace{-0.30cm}
\end{figure}

We now evaluate the performance of the protocols when both the client load and contention increases.
This experiment, reported in \reffig{fig:increasing_load}, runs over $5$ sites.
It employs a growing number of clients per site (from 32 to 20K), where each client submits commands with a payload of 4KB.
The top scenario of \reffig{fig:increasing_load} uses the same conflict rate as
in the previous experiments (2\%), while the bottom one uses a moderate conflict rate of 10\%.
The heatmap shows the hardware utilization (CPU, inbound and outbound network
bandwidth) for the case when the conflict rate is 2\%. %
For leaderless protocols, we measure the hardware utilization averaged across
all sites, whereas for FPaxos, we only show this measure at the leader site.
The experiment runs on a local cluster with emulated wide-area latencies, to have a full control over the hardware.

As seen in \reffig{fig:increasing_load}, the leader in FPaxos quickly becomes a
bottleneck when the load increases since it has to broadcast each command to all the processes.
For this reason, FPaxos provides the maximum throughput of only 53K ops/s with
$f = 1$ and of 45K ops/s with $f = 2$.
The protocol saturates at around 4K clients per site, when the outgoing network bandwidth at the leader
reaches 95\% usage.
The fact that the leader can be a bottleneck in leader-based protocol has
been reported by several prior works~\cite{atlas,epaxos,hermes,hovercraft,bpaxos}.

FPaxos is not affected by contention and the protocol has identical behavior for the two conflict rates.
On the contrary, Atlas performance degrades when contention increases.
With a low conflict rate (2\%), the protocol provides the maximum throughput of
129K ops/s with $f = 1$ and of 127K ops/s with $f = 2$.
As observed in the heatmap (bottom of \reffig{fig:increasing_load}), Atlas cannot fully leverage the available hardware.
CPU usage reaches at most 59\%, while network utilization reaches 41\%.
This low value is due to a bottleneck in the execution mechanism: its implementation, which follows the one by the authors of EPaxos, is single-threaded.
Increasing the conflict rate to 10\% further decreases hardware utilization:
the maximum CPU usage decreases to 40\% and network to 27\% (omitted from \reffig{fig:increasing_load}).
This sharp decrease is due to the dependency chains, whose sizes increase with higher contention, thus requiring fewer clients to bottleneck execution.
As a consequence, the throughput of Atlas decreases by 36\% with $f = 1$ (83K ops/s)
and by 48\% with $f = 2$ (67K ops/s).
As before, EPaxos performance (omitted from \reffig{fig:increasing_load}) lies between Atlas $f = 1$ and $f = 2$.

As we mentioned in \refsec{sec:sys_comparison}, Caesar exhibits inefficiencies even in its commit protocol.
For this reason, in \reffig{fig:increasing_load} we study the performance of Caesar in an ideal scenario where commands are executed as soon as they are committed.
Caesar's performance is capped respectively at 104K ops/s with 2\% conflicts and 32K ops/s with 10\% conflicts.
This performance decrease is due to Caesar's blocking mechanism (\refsec{sec:sys_comparison}) and is in line with the results reported in~\cite{caesar}.

\SYS delivers the maximum throughput of 230K ops/s.
This value is independent of the conflict rate and fault-tolerance level (i.e., $f \in \{1, 2\}$).
Moreover, it is \textsf{4.3-5.1x} better than FPaxos and \textsf{1.8-3.4x} better than Atlas.
Saturation occurs with 16K clients per site, when the CPU usage reaches 95\%.
At this point, network utilization is roughly equal to 80\%.
Latency in the protocol is almost unaffected until saturation.

\subsubsection*{Batching}
\begin{figure}[t]
    \vspace{-0.8em}
    \plot{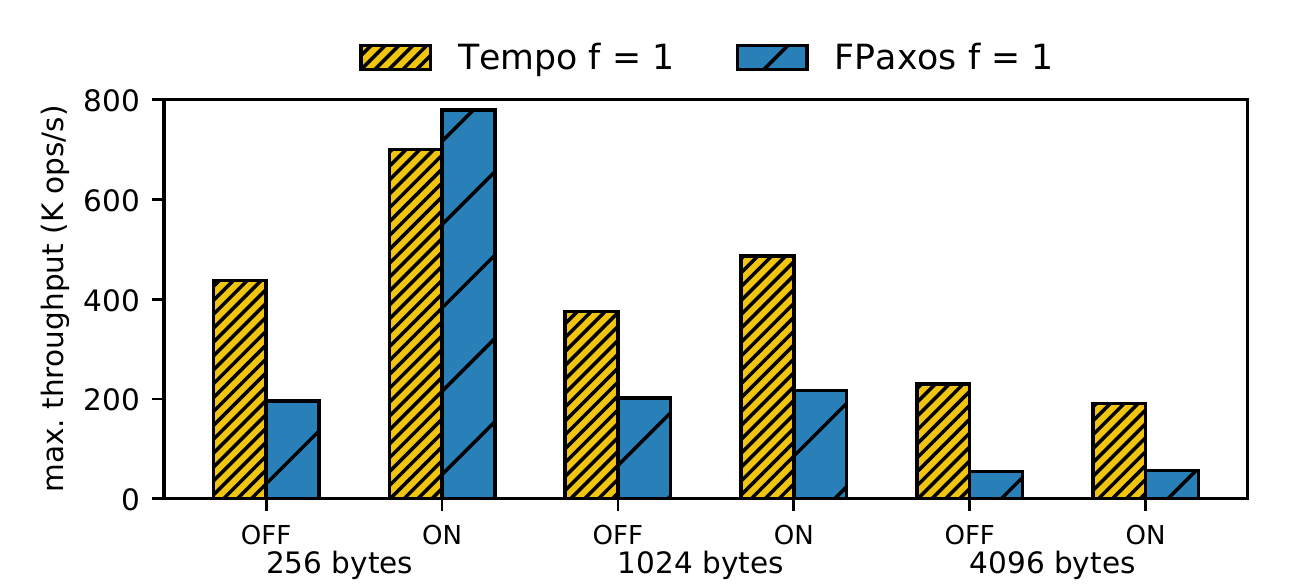}
    \caption{
        Maximum throughput with batching disabled (OFF) and enabled (ON)
        for 256, 1024 and 4096 bytes.
    }
    \label{fig:batching}
    \Description{TODO}
    \vspace{-0.30cm}
\end{figure}

We now compare the effects of batching in leader-based and leaderless protocols.
\reffig{fig:batching} depicts the maximum throughput of FPaxos and \SYS with batching disabled and enabled.
In this experiment, a batch is created at a site after 5ms or once $10^5$ commands are buffered, whichever is earlier.
Thus, each batch consists of several single-partition commands aggregated into one multi-partition command.
We consider 3 payload sizes: 256B, 1KB and 4KB.
The numbers for 4KB with batching disabled correspond to the ones in \reffig{fig:increasing_load}.
Because with 4KB and 1KB FPaxos bottlenecks in the network (\reffig{fig:increasing_load}), enabling batching does not help.
When the payload size is reduced further to 256B, the bottleneck shifts to the leader thread.
In this case, enabling batching allows FPaxos to increase its performance by \speedup{4}.
Since \SYS performs heavier computations than FPaxos, the use of batches in \SYS only brings a moderate improvement: \speedup{1.6} with 256B and \speedup{1.3} with 1KB.
In the worst case, with 4KB, the protocol can even perform less efficiently.

While batching can boost leader-based SMR protocols, the benefits are limited for leaderless ones. However, because leaderless protocols already efficiently balance resource usage across replicas, they can match or even outperform the performance of leader-based protocols, as seen in \reffig{fig:batching}.

\begin{figure}[t]
    \plot{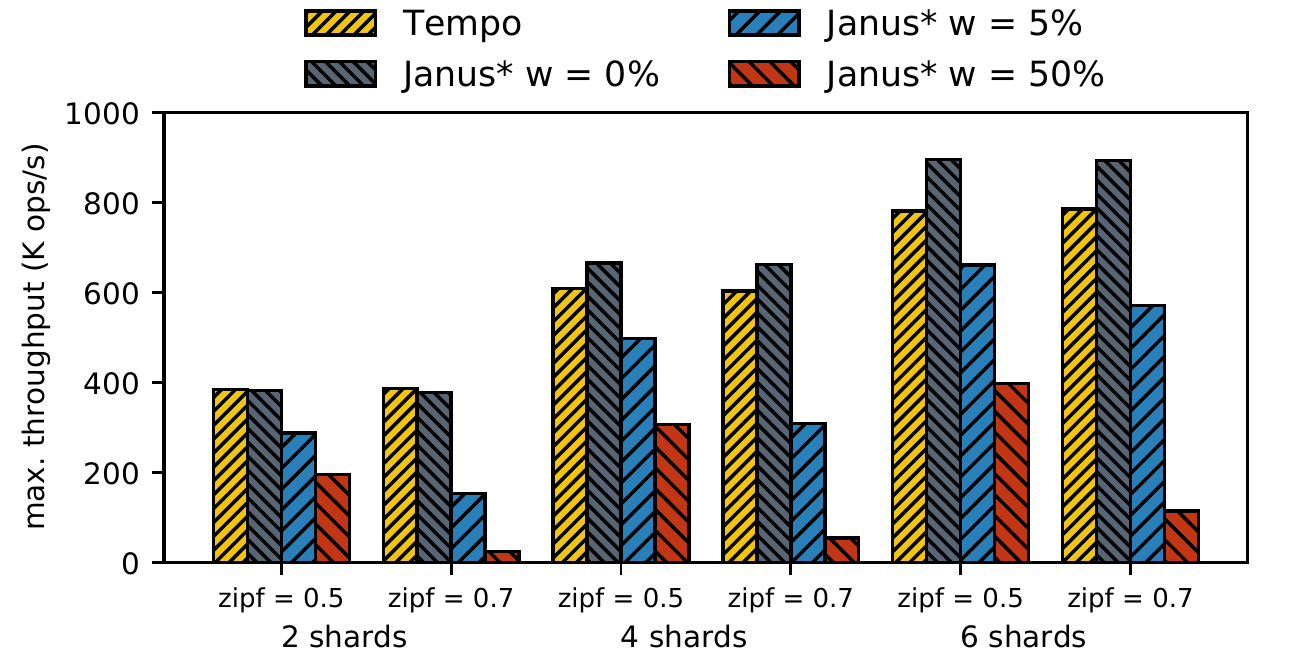}
    \caption{
        Maximum throughput with 3 sites per shard under low
        (\textsf{zipf = 0.5}) and moderate contention (\textsf{zipf = 0.7}).
        Three workloads are considered for Janus*: 0\% writes as the best-case
        scenario, 5\% writes and 50\% writes.
    }
    \label{fig:partial}
    \Description{TODO}
    \vspace{-0.30cm}
\end{figure}

\subsection{Partial Replication Deployment}
\label{sec:evaluation_partial}

We now compare \SYS with Janus* using the YCSB+T benchmark.
We define a \emph{shard} as set of several partitions co-located in the same machine.
Each partition contains a single YCSB key.
Each shard holds 1M keys and is replicated at $3$ sites (Ireland, N. California and Singapore)
emulated in our cluster.
Clients submit commands that access two keys picked at random following the YCSB access pattern (a zipfian distribution).
In \reffig{fig:partial} we show the maximum throughput for both \SYS and Janus* under low (\textsf{zipf = 0.5}) and moderate contention (\textsf{zipf = 0.7}).
For Janus*, we consider 3 YCSB workloads that vary the percentage of write commands (denoted by \textsf{w}):
read-only (\textsf{w = 0\%}, YCSB workload C),
read-heavy (\textsf{w = 5\%}, YCSB workload B), and
update-heavy (\textsf{w = 50\%}, YCSB workload A).
The read-only workload is a rare workload in SMR deployments.
It represents the best-case scenario for Janus*, which we use as a baseline.
Since \SYS does not distinguish between reads and writes (\refsec{sec:sys_comparison}), we have a single workload for this protocol.

Janus* performance is greatly affected by the ratio of writes and by contention.
More writes and higher contention translate into larger dependency sets, which bottleneck execution faster.
This is aggravated by the fact that Janus* is non-genuine, and thus requires cross-shard messages to order commands.
With \textsf{zipf = 0.5}, increasing \textsf{w} from 0\% to 5\% reduces throughput by 25-26\%.
Increasing \textsf{w} from 0\% to 50\% reduces throughput by 49-56\%.
When contention increases (\textsf{zipf = 0.7}), the above reductions on throughput are larger, reaching
36-60\% and
87\%-94\%, respectively.

\SYS provides nearly the same throughput as the best-case scenario for Janus* (\textsf{w = 0\%}).
Moreover, its performance is virtually unaffected by the increased contention.
This comes from the parallel and genuine execution brought by the use of timestamp stability (\refsec{sec:sys_partial}).
Overall, \SYS provides 385K ops/s with 2 shards, 606K ops/s with 4 shards, and
784K ops/s with 6 shards
(averaged over the two \textsf{zipf} values).
Compared to Janus* \textsf{w = 5\%} and Janus* \textsf{w = 50\%},
this represents respectively a speedup of \textsf{1.2-2.5x}
and \textsf{2-16x}.

The tail latency issues demonstrated in
\reffig{fig:tail_latency} also carry over to partial replication.
For example, with 6 shards, \textsf{zipf = 0.7} and $\af{w} = 5\%$,
the 99.99th percentile for Janus* reaches 1.3s, while \SYS provides 421ms.
We also ran the same set of workloads for the full replication case
and the speed up of \SYS with respect to EPaxos and Atlas is similar.

\section{Related Work}
\label{sec:related}

Timestamping (aka sequencing) is widely used in distributed systems.
In particular, many storage systems orchestrate data access using a fault-tolerant timestamping service~\cite{corfu,paxosstore, tango, percolator, calvin}, usually
implemented by a leader-based SMR protocol~\cite{paxos,raft}.
As reported in prior works, the leader is a potential bottleneck and is unfair with respect to client locations~\cite{atlas,epaxos,hermes,hovercraft,bpaxos}.
To sidestep these problems, leaderless protocols order commands in a fully decentralized manner.
Early protocols in this category, such as Mencius~\cite{mencius}, rotated the role of leader among processes.
However, this made the system run at the speed of the slowest replica.
More recent ones, such as EPaxos~\cite{epaxos} and its follow-ups \cite{atlas,bpaxos,gryff}, order commands by agreeing on a graph of dependencies (\refsec{sec:sys_comparison}).
\SYS builds on one of these follow-ups, Atlas~\cite{atlas}, which leverages the observation that correlated failures in geo-distributed systems are rare~\cite{spanner} to reduce the quorum size in leaderless SMR.
As demonstrated by our evaluation (\refsec{sec:evaluation_full}), dependency-based leaderless protocols exhibit high tail latency and suffer from bottlenecks due to their expensive execution mechanism.

Timestamping has been used in two previous leaderless SMR protocols.
Caesar~\cite{caesar}, which we discussed in \refsec{sec:sys_comparison} and \refsec{sec:evaluation}, suffers from similar problems to EPaxos.
Clock-RSM \cite{clockrsm} timestamps each newly submitted command with the coordinator's clock, and then records the association at $f+1$ processes using consensus.
Stability occurs when all the processes indicate that their clocks have passed the command's timestamp.
As a consequence, the protocol cannot transparently mask failures, like \SYS; these have to be handled via reconfiguration.
Its performance is also capped by the speed of the slowest replica, similarly to Mencius~\cite{mencius}.

Partial replication is a common way of scaling services that do not fit on a single machine.
Some partially replicated systems use a central node to manage access to data, made fault-tolerant via standard SMR techniques~\cite{gfs}.
Spanner~\cite{spanner} replaces the central node by a distributed protocol that layers two-phase commit on top of Paxos.
Granola~\cite{granola} follows a similar schema using Viewstamped Replication \cite{viewstamped}.
Other approaches rely on atomic multicast, a primitive ensuring the consistent delivery of messages across arbitrary groups of processes~\cite{Guerraoui2001,pstore}.
Atomic multicast can be seen as a special case of PSMR as defined in \refsec{sec:model}.

Janus \cite{janus} generalizes EPaxos to the setting of partial replication.
Its authors shows that for a large class of applications that require only one-shot transactions, Janus improves upon prior techniques, including MDCC~\cite{mdcc}, Tapir~\cite{tapir} and 2PC over Paxos~\cite{spanner}.
Our experiments demonstrate that \SYS significantly outperforms Janus due to its use of timestamps instead of explicit dependencies.
Unlike Janus, \SYS is also genuine, which translates into better performance.

\section{Conclusion}
\label{sec:conclusion}

We have presented \SYS{} -- a new SMR protocol for geo-distributed systems.
\SYS follows a leaderless approach, ordering commands in a fully decentralized
manner and thus offering similar quality of service to all clients. In contrast
to previous leaderless protocols, \SYS determines the order of command
execution solely based on scalar timestamps, and cleanly separates timestamp
assignment from detecting timestamp stability.
Moreover,
this mechanism easily extends to partial replication.
As shown in our evaluation, \SYS's approach enables the protocol to offer low tail latency and high throughput even under contended workloads.

\smallskip

\subsubsection*{Acknowledgments.}

We thank our shepherd, Natacha Crooks, as well as Antonios Katsarakis, Ricardo Macedo, Georges Younes for comments and suggestions.
We also thank Balaji Arun, Roberto Palmieri and Sebastiano Peluso for discussions about Caesar.
Vitor Enes was supported by an FCT
PhD Fellowship (PD/BD/142927/2018).
Pierre Sutra was supported by EU H2020 grant No 825184 and ANR grant 16-CE25-0013-04.
Alexey Gotsman was supported by an ERC Starting Grant RACCOON.
This work was partially supported by the AWS Cloud Credit for Research program.

\bibliographystyle{ACM-Reference-Format}
\bibliography{bib}

\iflong
  \clearpage
  \appendix
  \onecolumn

\section{Latency Data}
\label{sec:app_evaluation}

\reftab{tab:latencies} shows the average ping latencies between the 5 EC2 sites used in our evaluation (\refsec{sec:evaluation}):
Ireland (\textsf{eu-west-1}),
Northern California (\textsf{us-west-1}),
Singapore (\textsf{ap-southeast-1}),
Canada (\textsf{ca-central-1}),
and São Paulo (\textsf{sa-east-1}).

\begin{table}[h]
    \caption{
        Ping latency (milliseconds) between sites.
    }
    \label{tab:latencies}
    \vspace{-0.20cm}

    \setlength\tabcolsep{2.5pt}
    \setlength\aboverulesep{0pt}
    \setlength\belowrulesep{0pt}

    \begin{center}
        \begin{tabular}{ccccc}
            \toprule
                          & N. California & Singapore & Canada & S. Paulo \\
            \midrule
            Ireland       & 141           & 186       & 72     & 183      \\
            N. California &               & 181       & 78     & 190      \\
            Singapore     &               &           & 221    & 338      \\
            Canada        &               &           &        & 123      \\
            \toprule
        \end{tabular}
    \end{center}
    \vspace{-0.30cm}
\end{table}

\section{The Full \SYS Protocol}
\label{sec:app_sys}

Algorithms~\ref{algo:full} and~\ref{algo:full-liveness-and-execution} specify the full multi-partition \SYS protocol, including the liveness mechanisms omitted from \S\ref{sec:sys_recovery}.
The latter are described in the \refsec{sec:app_sys_liveness}.
\reftab{tab:vars} summarizes the data maintained by each process, where $\objs$ denotes the set of all partitions.
In the following, line numbers refer to the pseudocode in Algorithms~\ref{algo:full} and~\ref{algo:full-liveness-and-execution}.

\subsubsection*{\textbf{An additional optimization.}}
Notice that if a process in the fast quorum is slow, \SYS needs to execute recovery.
We can avoid this by having the processes receiving an $\MPayload$ message (\refline{algo:full:mpayload-recv}) also generate timestamp proposals and send them to the coordinator.
If the fast path quorum is slow to answer, these additional replies can be used to take the slow path, as long as a majority of processes replies (\refprop{prop:bump-at-majority}).

\begin{table}[h]
    \caption{\SYS variables at a process from partition $p$.}
    \label{tab:vars}
    \vspace{-0.20cm}

    \begin{center}
        \begin{tabular}{r@{}c@{}l@{}ll}
            \toprule
            $\cmd[id]$      & ${} \leftarrow {}$ & $\bot$        & ${} \in \cmds$                                      & Command
            \\
            $\ts[id]$       & ${} \leftarrow {}$ & $0$           & ${} \in \nat$                                       & Timestamp
            \\
            $\phase[id]$    & ${} \leftarrow {}$ & $\START$      &                                                     & Phase
            \\
            $\quorums[id]$   & ${} \leftarrow {}$ & $\varnothing$ & ${} \in \objs \hookrightarrow \pow{\procs_p}$ & Fast quorum used per partition
            \\
            $\bal[id]$      & ${} \leftarrow {}$ & $0$           & ${} \in \nat$                                       & Current ballot
            \\
            $\abal[id]$     & ${} \leftarrow {}$ & $0$           & ${} \in \nat$                                       & Last accepted ballot
            \\
            \midrule
            $\Clock$        & ${} \leftarrow {}$ & $0$           & ${} \in \nat$                                       & Current clock
            \\
            $\Detached$     & ${} \leftarrow {}$ & $\varnothing$ & ${} \in \pow{\procs_p \times \nat}$                 & Detached promises
            \\
            $\Attached[id]$ & ${} \leftarrow {}$ & $\varnothing$ & ${} \in \pow{\procs_p \times \nat}$                 & Attached promises
            \\
            $\Promises$     & ${} \leftarrow {}$ & $\varnothing$ & ${} \in \pow{\procs_p \times \nat}$                 & Known promises
            \\
            \toprule
        \end{tabular}
    \end{center}
\end{table}

\subsection{Liveness Protocol}
\label{sec:app_sys_liveness}

\SYS uses $\Omega$, the leader election failure detector~\cite{weakest-fd}, which ensures that from some point on, all correct processes nominate the same correct process as the leader.
\SYS runs an instance of $\Omega$ per partition $p$.
In \refalgo{algo:full-liveness-and-execution}, $\leader_p$ denotes the current leader nominated for partition $p$ at process $i$.
We say that $\leader_p$ stabilizes when it stops changing at all correct processes in $p$.
\SYS also uses $\procs_{c}^i$, which we call the \emph{partition covering} failure detector (\refsec{sec:sys_partial}).
At each process $i$ and for every command $c$, $\procs_{c}^i$ returns a set of processes $J$ such that,
for every partition $p$ accessed by $c$, $J$ contains one process close to $i$ that replicates $p$.
Eventually, $\procs_{c}^i$ only returns correct processes.
In the definition above, the closeness between replicas is measured is in terms of latency.
Returning close replicas is needed for performance but not necessary for the liveness of the protocol.
Both $\procs_{c}^i$ and $\Omega$ are easily implementable under our assumption of eventual synchrony (\refsec{sec:model}).

For every command $\id$ with $\id \in \pending$ (\refline{algo:full:liveness}), process $i$ is allowed to invoke $\FunRecover(\id)$ at \refline{algo:full:liveness-recover-pre} only if it is the leader of partition $p$ according to $\leader_p$.
Furthermore, it only invokes $\FunRecover(\id)$ at \refline{algo:full:liveness-recover-pre} if it has not yet participated in consensus (i.e., $\bal[\id] = 0$) or, if it did, the consensus was lead by another process (i.e., $\FunBalLeader(\bal[\id]) \not = i$).
In particular, process $i$ does not invoke $\FunRecover(\id)$ at \refline{algo:full:liveness-recover-pre} if it is the leader of $\bal[\id]$ (i.e., if $\FunBalLeader(\bal[\id]) = i$).
This ensures that process $i$ does disrupt a recovery lead by itself.

\renewcommand{\SubAlgo}[2]{#1 \SubAlgoBlock{#2}\vspace{0.1cm}}

\begin{algorithm}[!hp]
  \setcounter{AlgoLine}{0}
  \vspace{-1em}
  \begin{multicols}{2}
    \SubAlgo{$\FunSubmit(c)$ \label{algo:full:submit}}{
      \Pre $i \in \procs_c$ \label{algo:full:submit-pre} \;
      $\id \leftarrow \af{next\_id}()$;
      $\qs \leftarrow \af{fast\_quorums}(i, \procs_c)$ \;
      \Send $\MSubmit(\id, c, \qs)$ \To $\procs_c^i$
    }

    \SubAlgo{\Receive $\MSubmit(\id, c, \qs)$ \label{algo:full:msubmit-recv}}{
      $t \leftarrow \Clock + 1$ \label{algo:full:msubmit-proposal} \;
      \Send $\MPropose(\id, c, \qs, t)$ \To $\qs[p]$ \label{algo:full:mpropose-send} \;
      \Send $\MPayload(\id, c, \qs)$ \To $\procs_p \setminus \qs[p]$ \label{algo:full:mpayload-send} \;
    }

    \SubAlgo{\Receive $\MPayload(\id, c, \qs)$\label{algo:full:mpayload-recv}}{
      \Pre $\id \in \start$ \label{algo:full:mpayload-pre} \;
      $\cmd[\id] \leftarrow c$;
      $\quorums[\id] \leftarrow \qs$;
      $\phase[\id] \leftarrow \PAYLOAD$ \label{algo:full:payload-phase} \!
    }

    \SubAlgo{\Receive $\MPropose(\id, c, \qs, t)$ \From $j$ \label{algo:full:mpropose-recv}}{
      \Pre $\id \in \start$ \label{algo:full:mpropose-pre} \;
      $\cmd[\id] \leftarrow c$;
      $\quorums[\id] \leftarrow \qs$;
      $\phase[\id] \leftarrow \PROPOSE$ \label{algo:full:propose-phase} \;
      $\ts[\id] \leftarrow \FunTs(\id, t)$ \label{algo:full:mpropose-proposal} \;
      \Send $\MProposeAck(\id, \ts[\id])$ \To $j$ \label{algo:full:mpropose-ack-send} \;
      \Send $\MBump(\id, \ts[\id])$ \To $\procs_c^i$ \label{algo:full:mbump-send}
    }

    \SubAlgo{\Receive $\MBump(\id, t)$ \label{algo:full:mbump-recv}}{
      \Pre $\id \in \propose$ \;
      $\FunBump(t)$ \label{algo:full:mbump-bump}
    }

    \SubAlgo{\Receive $\MProposeAck(\id, t_j)$ \From $\forall j \in Q$ \label{algo:full:mpropose-ack-recv}}{
      \Pre $\id \in \propose \land Q = \quorums[\id][p]$ \label{algo:full:mpropose-ack-pre} \;
      $t \leftarrow \max \{ t_j \mid j \in Q \}$ \label{algo:full:mpropose-ack-max} \;
      \lIf{$\af{count}(t) \geq f$\label{algo:full:fast-path-condition}}{%
        \Send $\MCommit(\id, t)$ \To $\procs_{\cmd[\id]}$ \label{algo:full:fast-path}
      }
      \lElse{%
        \Send $\MConsensus(\id, t, i)$ \To $\procs_p$ \label{algo:full:slow-path}
      }
    }

    \SubAlgo{\Receive $\MCommit(\id, t_j)$ \From $j \in \procs_{\cmd[\id]}^i$\label{algo:full:mcommit-recv}}{
      \Pre $\id \in \pending$ \;
      $\ts[\id] \leftarrow \max \{ t_j \mid j \in P \}$;
      $\phase[\id] \leftarrow \COMMIT$ \label{algo:full:final-ts}\label{algo:full:commit-phase} \;
      $\FunBump(\ts[\id])$ \label{algo:full:mcommit-bump}
    }

    \SubAlgo{\Receive $\MConsensus(\id, t, b)$ \From $j$}{
      \Pre $\bal[\id] \leq b$ \label{algo:full:mconsensus-pre} \;
      $\ts[\id] \leftarrow t$;
      $\bal[\id] \leftarrow b$;
      $\abal[\id] \leftarrow b$ \;
      $\FunBump(t)$ \label{algo:full:mconsensus-bump} \;
      \Send $\MConsensusAck(\id, b)$ \To $j$
    }

    \SubAlgo{\Receive $\MConsensusAck(\id, b)$ \From $Q$\label{algo:full:receive-mconsensus}}{
      \Pre $\bal[\id] = b \land \setsize{Q} = f + 1$ \label{algo:full:mconsensus-ack-pre} \;
      \Send $\MCommit(\id, \ts[\id])$ \To $\procs_{\cmd[\id]}$ \label{algo:full:consensus-end}
    }

    \columnbreak

    \SubAlgo{$\FunRecover(\id)$ \label{algo:full:fun-recover}}{
      \Pre $\id \in \pending$  \label{algo:full:fun-recover-pre} \;
      $b \leftarrow i + r (\floor{\frac{\bal[\id] - 1}{r}} + 1)$ \label{algo:full:my-ballot} \;
      \Send $\MRec(\id, b)$ \To $\procs_p$ \label{algo:full:mrec-send} \;
    }

    \SubAlgo{\mbox{\Receive $\MRec(\id, b)$ \From $j$\label{algo:full:mrec-recv}}}{
      \Pre $\id \in \pending \land \bal[\id] < b$ \label{algo:full:mrec-pre} \;

      \If{$\bal[\id] = 0$ \label{algo:full:mrec-first-see}}{
        \If{$\phase[\id] = \PAYLOAD$ \label{algo:full:mrec-at-payload}}{
          $\ts[\id] \leftarrow \FunTs(\id, 0)$ \label{algo:full:mrec-proposal} \;
          $\phase[\id] \leftarrow \RECOVERR$ \label{algo:full:recoverr-phase}
        }
        \ElseIf{$\phase[\id] = \PROPOSE$ \label{algo:full:mrec-at-propose}}{
          $\phase[\id] \leftarrow \RECOVERP$ \label{algo:full:recoverp-phase}
        }
      }
      $\bal[\id] \leftarrow b$ \label{algo:full:mrec-new-bal} \;
      \Send $\MRecAck(\id, \ts[\id], \phase[\id], \abal[\id], b)$ \To $j$
    }

    \SubAlgo{\Receive $\MRecAck(\id, t_j, \ph_j, \ab_j, b)$ \From $\forall j \in Q$\label{algo:full:mrec-ack-recv}}{
      \Pre $\bal[\id] = b \land \setsize{Q} = r - f$ \label{algo:full:mrec-ack-pre} \;
      \If{$\exists k \in Q \cdot \ab_k \not = 0$ \label{algo:full:previously-accepted}}{
        \Let $k$ be such that $\ab_k$ is maximal \label{algo:full:highest-ballot} \;
        \Send $\MConsensus(\id, t_k, b)$ \To $\procs_p$ \label{algo:full:highest-ballot-mconsensus}
      }
      \Else{
        $I \leftarrow Q \cap \quorums[\id][p]$ \label{algo:full:intersection} \;
        $s \leftarrow \initial_p(\id) \in I \lor \exists k \in I \cdot \ph_k = \RECOVERR$ \label{algo:full:coordinator-check} \;
        $Q' \leftarrow $ \leIf{$s$}{$Q$}{$I$} \label{algo:full:quorum-selection}
        $t \leftarrow \max \{ t_j \mid j \in Q' \}$ \label{algo:full:rec-max} \;
        \Send $\MConsensus(\id, t, b)$ \To $\procs_p$ \label{algo:full:first-rec-mconsensus}
      }
    }

    \algrule[0.5pt]

    \SubAlgo{$\FunTs(\id, m)$ \label{algo:full:fun-ts}}{
      $t \leftarrow \max(m, \Clock + 1)$ \label{algo:full:timestamp-max} \;
      \mbox{$\Detached \leftarrow \Detached \,{\union}\, \{ \tup{i, u} \,{\mid}\, \Clock \,{+}\, 1 \,{\leq}\, u \,{\leq}\, t\,{-}\,1 \}$}%
      \label{algo:full:detached} \\
      $\Attached[\id] \leftarrow \{ \tup{i, t} \}$ \label{algo:full:attached} \;
      $\Clock \leftarrow t$ \label{algo:full:fun-ts-bump} \;
      \Return $t$ \label{algo:full:fun-ts-return}
    }

    \SubAlgo{$\FunBump(t)$ \label{algo:full:fun-bump}}{
      $t \leftarrow \max(t, \Clock)$ \;
      \mbox{$\Detached \leftarrow \Detached \union \{ \tup{i, u} \mid \Clock \,{+}\, 1 \leq u \leq t \}$} \\
      $\Clock \leftarrow t$
    }

    \SubAlgo{$\FunBalLeader(b)$} {
      \Return $b - r * \bigfloor{\frac{b - 1}{r}}$
    }
  \end{multicols}

  \newcounter{AlgoFullLastLine}
  \setcounter{AlgoFullLastLine}{\value{AlgoLine}}

  \caption{
    \SYS commit and recovery protocol at process $i \in \procs_p$.
  }
  \label{algo:full}
\end{algorithm}

\FloatBarrier
\begin{algorithm}[h]
  \setcounter{AlgoLine}{\value{AlgoFullLastLine}}

  \vspace{-1em}
  \begin{multicols}{2}

    \SubAlgo{\Periodically \label{algo:full:liveness}}{
      \For{$\id \in \pending$ \label{algo:full:liveness-pending}}{
        \Send $\MPayload(\id, \cmd[\id], \quorums[\id])$ \To $\procs_{\cmd[\id]}$ \label{algo:full:liveness-mpayload-send} \;
        \lIf{$\leader_p = i \land (\bal[\id] = 0 \lor \FunBalLeader(\bal[\id]) \not= i)$ \label{algo:full:liveness-recover-pre}}{%
          $\FunRecover(\id)$ \label{algo:full:liveness-recover}
        }
      }
    }

    \SubAlgo{\mbox{\Receive $\MConsensus(\id, \_, b)$ or $\MRec(\id, b)$ \From $j$ \label{algo:full:mconsensus-or-mrec-recv}}}{
      \Pre $\bal[\id] > b$ \label{algo:full:mconsensus-or-mrec-pre-high-ballot} \;
      \Send $\MRecNAck(\id, \bal[\id])$ \To $j$ \label{algo:full:mrecnack-send}
    }

    \SubAlgo{\Receive $\MRecNAck(\id, b)$}{
      \Pre $\leader_p = i \land \bal[\id] < b$ \label{algo:full:mrecnack-pre} \;
      $\bal[\id] \leftarrow b$ \label{algo:full:mrecnack-join-ballot} \;
      $\FunRecover(\id)$ \label{algo:full:mrecnack-recover}
    }

    \SubAlgo{\Receive $\MCommitRequest(\id)$ \From $j$ \label{algo:full:mcommit-request-recv}}{
      \Pre $\id \in \commit \union \execute$ \label{algo:full:mcommit-request-pre} \;
      \Send $\MPayload(\id, \cmd[\id], \quorums[\id])$ \To $j$ \label{algo:full:mcommit-request-mpayload-send} \;
      \Send $\MCommit(\id, \ts[\id])$ \To $j$ \label{algo:full:mcommit-request-mcommit-send} \;
    }

    \SubAlgo{\Periodically}{
      \Send $\MPromises(\Detached, \Attached)$ \To $\procs_p$ \label{algo:full:mvotes-send}
    }

    \SubAlgo{\Receive $\MPromises(D, A)$ \From $j$ \label{algo:full:mvotes-recv}}{
      $C \leftarrow \bigunion \{ a \mid \tup{\id, a} \in A \land \id \in \commit \union \execute \}$ \label{algo:full:learn-attached} \;
      $\Promises \leftarrow \Promises \union D \union C$ \label{algo:full:learn-allowed-votes} \;
      \For{$\tup{\id, \_} \in A \cdot \id \not\in \commit \union \execute$ \label{algo:full:attached-not-committed} }{
        \Send $\MCommitRequest(\id)$ \To $\procs_{\cmd[\id]}$ \label{algo:full:mcommit-request-send}
      }
    }

    \SubAlgo{\Periodically\label{algo:full:executehandler}}{
      \mbox{$h \leftarrow \sort \{ \FunHCV(j) \mid j \in \procs_p \}$} \label{algo:full:sort} \;
      $\mathit{ids} \leftarrow \{ \id \in \commit \mid \ts[\id] \leq h[\hspace{1pt}\floor{\frac{r}{2}}\hspace{1pt}] \}$ \label{algo:full:stable-and-committed} \;
      \For{$\id \in \mathit{ids}$ \OrderedBy $\tup{\ts[\id], \id}$ \label{algo:full:order}}{
        \Send $\MStable(\id)$ \To $\procs_{\cmd[\id]}$ \label{algo:full:stable-send} \;
        \mbox{\textbf{wait} \Receive $\MStable(\id)$ \From $\forall j \in \procs_{\cmd[\id]}^i$} \label{algo:full:stable-recv} \;
        $\exec_p(\cmd[\id])$;
        $\phase[\id] \leftarrow \EXECUTE$ \label{algo:full:execute}\label{algo:full:execute-phase}
      }
    }

    \SubAlgo{$\FunHCV (j)$}{
      $\max\{c \in \nat \mid \forall u \in \{1 \dots c \} \cdot \tup{j, u} \in \Promises \}$ \label{algo:full:votes}
    }

  \end{multicols}

  \caption{
    \SYS liveness and execution protocol at process $i \in \procs_p$.
  }
  \label{algo:full-liveness-and-execution}
\end{algorithm}

For a leader to make progress with some $\MRec(\id, b)$ message, it is required that $n - f$ processes (\refline{algo:full:mrec-ack-pre}) have their $\bal[\id] < b$ (\refline{algo:full:mrec-pre}).
This may not always be the case because before the variable $\leader_p$ stabilizes, any process can invoke $\FunRecover(\id)$ at \refline{algo:full:liveness-recover-pre} if it thinks it is the leader.
To help the leader select a high enough ballot, and thus ensure it will make progress, we introduce a new message type, $\MRecNAck$.
A process sends an $\MRecNAck(\id, \bal[\id])$ at \refline{algo:full:mrecnack-send} when it receives an $\MConsensus(\id, \_, b)$ or $\MRec(\id, b)$ (\refline{algo:full:mconsensus-or-mrec-recv}) with some ballot number $b$ lower than its $\bal[\id]$ (\refline{algo:full:mconsensus-or-mrec-pre-high-ballot}).
When process $i$ receives an $\MRecNAck(\id, b)$ with some ballot number $b$ higher than its $\bal[\id]$, if it is still the leader (\refline{algo:full:mrecnack-pre}), it joins ballot $b$ (\refline{algo:full:mrecnack-join-ballot}) and invokes $\FunRecover(\id)$ again.
This results in process $i$ sending a new $\MRec$ with some ballot higher than $b$ lead by itself.
As only $\leader_p$ is allowed to invoke $\FunRecover(\id)$ at \refline{algo:full:liveness-recover-pre} and \refline{algo:full:mrecnack-pre}, and since $\leader_p$ eventually stabilizes, this mechanism ensures that eventually the stable leader will start a high enough ballot in which enough processes will participate.

Given a command $c$ with identifier $\id$ submitted by a correct process, a correct process in $\procs_c$ can only commit $c$ if it has $\id \in \pending$ and receives an $\MCommit(\id, \_)$ from every partition accessed by $c$ (\refline{algo:full:mcommit-recv}).
Next, we let $p$ be one such partition, and we explain how every correct process in $\procs_c$ eventually has $\id \in \pending$ and receives such an $\MCommit(\id, \_)$ sent by some correct process in $\procs_p$.
We consider two distinct scenarios.

In the first scenario, some correct process $i \in \procs_p$ already has $\id \in \commit \union \execute$.
This scenario is addressed by adding two new lines to the $\MPromises$ handler: \refline{algo:full:attached-not-committed} and \refline{algo:full:mcommit-request-send}.
Since $\id \in \commit \union \execute$ at $i$, by
\refprop{prop:bump-at-majority} there is a majority of processes in each partition $q$ accessed by $c$ that have called $\FunTs(\id, \_)$, and have thus generated a promise attached to $\id$.
Moreover, given that at most $f$ processes can fail, at least one process from such a majority is correct.
Let one of these processes be $j \in \procs_q$.
Due to \refline{algo:full:mvotes-send}, process $j$ periodically sends an $\MPromises$ message that contains a promise attached to $\id$.
When a process $k \in \procs_q$ receives such a message, if $\id$ is neither committed nor executed locally (\refline{algo:full:attached-not-committed}), $k$ sends an $\MCommitRequest(\id)$ to $\procs_c$.
In particular, it sends such message to process $i \in \procs_p$.
As $\id \in \commit \union \execute$ at process $i$, when $i$ receives the $\MCommitRequest(\id)$ (\refline{algo:full:mcommit-request-recv}), it replies with an $\MPayload(\id, \_, \_)$ and an $\MCommit(\id, \_)$.
In this way, any process $k \in \procs_q$ will eventually have $\id \in \pending$ and receive an $\MCommit(\id, \_)$ sent by some correct process in $\procs_p$, as required.

In the second scenario, no correct process in $\procs_p$ has $\id \in \commit \union \execute$ but some correct process in $\procs_c$ has $\id \in \pending$.
Due to \refline{algo:full:liveness-mpayload-send}, such process sends an $\MPayload$ message to $\procs_c$.
Thus, every correct process in $\procs_p$ eventually has $\id \in \pending$.
In particular, this allows $\leader_p$ to invoke $\FunRecover(\id)$ and the processes in $\procs_p$ to react to the $\MRec(\id, \_)$ message by $\leader_p$.
Hence, after picking a high enough ballot using the mechanism described earlier, $\leader_p$ will eventually send an $\MCommit(\id, \_)$ to $\procs_c$, as required.

Finally, note that calling $\FunRecover(\id)$ for any command $\id$ such that $\id \in \pending$ (\refline{algo:full:liveness}) can disrupt the fast path.
This does not affect liveness but may degrade performance.
Thus, to attain good performance, any implementation of \SYS should only start calling $\FunRecover(\id)$ after some reasonable timeout on $\id$.
In order to save bandwidth, the sending of $\MCommitRequest$ messages can also be delayed in the hope that such information will be received anyway.

\section{Correctness}
\label{sec:app_correctness}

In this section we prove that the \SYS protocol satisfies the PSMR specification (\refsec{sec:model}).
We omit the trivial proof of Validity, and prove Ordering in \refsec{sec:correctness-safety} and Liveness in \refsec{sec:correctness-liveness}.

\subsection{Proof of Ordering}
\label{sec:correctness-safety}

Consider the auxiliary invariants below:
\begin{enumerate}
  \setcounter{enumi}{0}
  \item \label{inv:consensus-ballot}
        Assume $\MConsensus(\_, \_, b)$ has been sent by process $i$. Then $b = i$ or $b > r$.

  \item \label{inv:unique-consensus}
        Assume $\MConsensus(\id, t, b)$ and $\MConsensus(\id, t', b')$ have been sent.
        If $b = b'$, then $t = t'$.

  \item \label{inv:mrecack-ballots}
        Assume $\MRecAck(\_, \_,  \_, ab, b)$ has been sent by some process.
        Then $ab < b$.

  \item \label{inv:after-leader}
        Assume $\MConsensusAck(\id, b)$ and $\MRecAck(\id, \_,  \_, ab, b')$ have been sent by some process.
        If $b' > b$, then $b \le ab < b'$ and $ab \not = 0$.

  \item \label{inv:not-start-then-payload}
        If $\id \not\in \start$ at some process then the process knows $\cmd[\id]$ and $\quorums[\id]$.

  \item \label{inv:not-start-at-mrec-ack}
        If a process executes the $\MConsensusAck(\id, \_)$ or $\MRecAck(\id, \_, \_, \_, \_)$ handlers then $\id \not\in \start$.

  \item \label{inv:main-consensus}
        Assume a slow quorum has received $\MConsensus(\id, t, b)$ and responded to it with $\MConsensusAck(\id, b)$.
        For any $\MConsensus(\id, t', b')$ sent, if $b' > b$, then $t' = t$.

  \item \label{inv:main-fastpath}
        Assume $\MCommit(\id, t)$ has been sent at \refline{algo:full:fast-path}.
        Then for any $\MConsensus(\id, t', \_)$ sent, $t' = t$.
\end{enumerate}

Invariants~\ref{inv:consensus-ballot}-\ref{inv:not-start-at-mrec-ack} easily follow from the
structure of the protocol.
Next we prove the rest of the invariants.
Then we prove \refprop{prop:agreement} and \refprop{prop:bump-at-majority} (the latter is used in the proof of \reftheorem{theorem:stable}).
Finally, we introduce and prove two lemmas that are then used to prove the Ordering property of the PSMR specification.

\paragraph{\it Proof of Invariant~\ref{inv:main-consensus}.}
Assume that at some point
\begin{quote}
  (*) a slow quorum has received $\MConsensus(\id, t, b)$ and responded to it with $\MConsensusAck(\id, b)$.
\end{quote}
We prove by induction on $b'$ that,
if a process $i$ sends $\MConsensus(\id, t', b')$ with $b' > b$,
then $t' = t$.
Given some $b^*$, assume this property holds for all $b' < b^*$.
We now show that it holds for $b' = b^*$.
We make a case split depending on the transition of process $i$ that sends the
$\MConsensus$ message.

First, assume that process $i$ sends $\MConsensus$ at \refline{algo:full:slow-path}.
In this case, $b' = i$.
Since $b' > b$, we have $b < i$.
But this contradicts Invariant~\ref{inv:consensus-ballot}.
Hence, this case is impossible.

The remaining case is when process $i$ sends $\MConsensus$ during the transition at \refline{algo:full:mrec-ack-recv}.
In this case, $i$ has received
$$
  \MRecAck(\id, t_j, \_, \ab_j, b')
$$
from all processes $j$ in a recovery quorum $Q^R$.
Let $\abmax = \max \{\ab_j \mid j \in Q^R\}$; then by Invariant~\ref{inv:mrecack-ballots} we have $\abmax < b'$.

Since the recovery quorum $Q^R$ has size $r - f$ and the slow quorum from (*) has size $f + 1$,
we get that at least one process in $Q^R$ must have received the $\MConsensus(\id, t, b)$ message and responded to it with $\MConsensusAck(\id, b)$.
Let one of these processes be $l$.
Since $b' > b$,
by Invariant~\ref{inv:after-leader} we have $\ab_l \not = 0$, and thus
process $i$ executes \refline{algo:full:highest-ballot-mconsensus}.
By Invariant~\ref{inv:after-leader} we also have $b \le \ab_l$
and thus $b \le \abmax$.

Consider an arbitrary process $k \in Q^R$, selected at \refline{algo:full:highest-ballot},
such that $\ab_k = \abmax$. We now prove that $t_k = t$.
If $\abmax > b$, then since $\abmax < b'$,
by induction hypothesis we have $t_k = t$, as required.
If $\abmax = b$, then
since $\abmax \not = 0$, process $k$ has received some
$\MConsensus(\id, \_, \abmax)$ message.
By Invariant~\ref{inv:unique-consensus},
process $k$ must have received the same $\MConsensus(\id, t, \abmax)$
received by process $l$.
Upon receiving this message, process $k$ stores $t$ in $\ts$ and does not change this value at \refline{algo:full:mrec-proposal}:
$\abmax \not = 0$ and thus $\bal[id]$ cannot be $0$ at \refline{algo:full:mrec-first-see}.
Thus process $k$ must have sent $\MRecAck(\id, t_k, \_, \abmax, b')$
with $t_k = t$, which concludes the proof.\qed

\paragraph{\it Proof of Invariant~\ref{inv:main-fastpath}.}
Assume $\MCommit(\id, t)$ has been sent at \refline{algo:full:fast-path}.
Then the process that sent this $\MCommit$ message must be process $\initial_p(\id)$.
Moreover, we have that for some fast quorum mapping $Q^F$ such that $\initial_p(\id) \in Q^F[p]$:
\begin{quote}
  (*) every process $j \in Q^F[p]$ has received $\MPropose(\id, c, Q^F, m)$ and responded with $\MProposeAck(\id, t_j)$ such that
  $t = \max \{ t_j \mid j \in Q^F[p] \}$.
\end{quote}
We prove by induction on $b$ that, if a process $i$ sends $\MConsensus(\id, t', b)$,
then $t' = t$.
Given some $b^*$, assume this property holds for all $b < b^*$.
We now show that it holds for $b = b^*$.

First note that process $i$ cannot send $\MConsensus$ at \refline{algo:full:slow-path},
since in this case we would have $i = \initial_p(\id)$,
and $\initial_p(\id)$ took the fast path at \refline{algo:full:fast-path}.
Hence, process $i$ must have sent $\MConsensus$ during the transition at \refline{algo:full:mrec-ack-recv}.
In this case, $i$ has received
$$
  \MRecAck(\id, t_j, \ph_j, \ab_j, b)
$$
from all processes $j$ in a recovery quorum $Q^R$.

If $\MConsensus$ is sent at \refline{algo:full:highest-ballot-mconsensus},
then we have $\ab_k > 0$ for the process $k \in Q^R$ selected at \refline{algo:full:highest-ballot}.
In this case, before sending $\MRecAck$, process $k$ must have received
$$
  \MConsensus(\id, t_k, \ab_k)
$$
with $\ab_k < b$.
Then by induction hypothesis we have $t' = t_k = t$.
This establishes the required.

If $\MConsensus$ is not sent in \refline{algo:full:highest-ballot-mconsensus},
then we have $\ab_k = 0$ for all processes $k \in Q^R$.
In this case, process $i$ sends $\MConsensus$ in
\refline{algo:full:first-rec-mconsensus}.
Since the recovery quorum $Q^R$ has size $r - f$ and the fast quorum $Q^F[p]$ from (*) has size $\floor{\frac{r}{2}} + f$, we have that
\begin{quote}
  (**) at least $\floor{\frac{r}{2}}$ processes in $Q^R$ are part of $Q^F[p]$ and thus must have received $\MPropose(\id, c, Q^F, m)$ and responded to it with $\MProposeAck$.
\end{quote}

Let $I$ be the set of processes $Q^R \cap Q^F[p]$ (\refline{algo:full:intersection}).
By our assumption, process $\initial_p(\id)$ sent an $\MCommit(\id, t)$ at \refline{algo:full:fast-path}, and thus $\id \in \commit \union \execute$ at this process.
Then due to the check $\id \in \pending$ at \refline{algo:full:mrec-pre}, this process did not reply to $\MRec$.
Hence, $\initial_p(\id)$ is not part of the recovery quorum, i.e., $\initial_p(\id) \not\in I$ at \refline{algo:full:coordinator-check}.
Moreover, since the initial coordinator takes the fast path at \refline{algo:full:fast-path},
all fast quorum processes have set $\phase[id]$ to $\PROPOSE$ when processing the $\MPropose$ from the coordinator (\refline{algo:full:propose-phase}).
Due to this and to the check at \refline{algo:full:mrec-at-propose}, their $\phase[\id]$ value is set to $\RECOVERP$ in \refline{algo:full:recoverp-phase},
and thus, we have that all fast quorum processes that replied report $\RECOVERP$ in their $\MRecAck$ message, i.e., $\forall k \in I \cdot \ph_k = \RECOVERP$ at \refline{algo:full:coordinator-check}.
It follows that the condition at \refline{algo:full:coordinator-check} does not hold and thus the quorum selected is $I$ (\refline{algo:full:quorum-selection}).
By \refprop{prop:recovery}, the fast path proposal
$t = \max \{ t_j \mid j \in Q^F[p] \}$
can be
obtained by selecting the highest proposal sent in
$\MPropose$ by any $\floor{\frac{r}{2}}$
fast-quorum processes (excluding the initial coordinator).
By (**), and since all processes $k \in I$ have $\ab_k = 0$,
then all processes in $I$ replied with the timestamp proposal that was sent to the initial coordinator.
Thus, by \refprop{prop:recovery} we have
$t = \max \{ t_j \mid j \in Q^F[p] \} = \max \{ t_j \mid j \in I \} = t'$,
which concludes the proof.
\qed

\paragraph{\it Proof of \refprop{prop:agreement}.}
Assume that $\MCommit(\id, t)$ and $\MCommit(\id, t')$ have been sent.
We prove that $t = t'$.

Note that, if an $\MCommit(\id, t)$ was sent at \refline{algo:full:mcommit-request-mcommit-send},
then some process sent an $\MCommit(\id, t)$ at \refline{algo:full:fast-path} or \refline{algo:full:consensus-end}.
Hence, without loss of generality,
we can assume that the two $\MCommit$ under consideration were sent at \refline{algo:full:fast-path} or at \refline{algo:full:consensus-end}.
We can also assume that the two $\MCommit$ have been sent by different processes.
Only one process can send an $\MCommit$ at \refline{algo:full:fast-path} and only once.
Hence, it is sufficient to only consider the following two cases.

Assume first that both $\MCommit$ messages are sent at \refline{algo:full:consensus-end}.
Then for some $b$, a slow quorum has received $\MConsensus(\id, t, b)$ and responded to it with $\MConsensusAck(\id, b)$.
Likewise, for some $b'$, a slow quorum has received $\MConsensus(\id, t', b')$ and responded to it with $\MConsensusAck(\id, b')$.
Assume without loss of generality that $b \leq b'$.
If $b < b'$, then $t' = t$ by Invariant~\ref{inv:main-consensus}.
If $b = b'$, then $t' = t$ by Invariant~\ref{inv:unique-consensus}.
Hence, in this case $t' = t$, as required.

Assume now that $\MCommit(\id, t)$ was sent at \refline{algo:full:fast-path} and $\MCommit(\id, t')$ at \refline{algo:full:consensus-end}.
Then for some $b$, a slow quorum has received $\MConsensus(\id, t', b)$ and responded to it with $\MConsensusAck(\id, b)$.
Then by Invariant~\ref{inv:main-fastpath}, we must have $t' = t$,
as required.
\qed

\paragraph{\it Proof of \refprop{prop:bump-at-majority}.}
Assume that $\MCommit(\id, t)$ has been sent.
We prove that there exists a quorum $\widehat{Q}$
with $|\widehat{Q}| \geq \floor{\frac{r}{2}} + 1$
and $\widehat{t}$ such that
$t = \max \{ \widehat{t}_j \mid j \in \widehat{Q} \}$,
where each process $j \in \widehat{Q}$ computes its $\widehat{t}_j$ in either
\refline{algo:full:mpropose-proposal} or \refline{algo:full:mrec-proposal}
using function $\FunTs$
and sends it in either
$\MProposeAck(\id, \widehat{t}_j)$ or $\MRecAck(\id, \widehat{t}_j, \_, 0, \_)$.

The computation of $t$ occurs either in the transition at
\refline{algo:full:mpropose-ack-max} or at \refline{algo:full:rec-max}.
If the computation of $t$ occurs at
\refline{algo:full:mpropose-ack-max},
then the quorum $Q$ defined at \refline{algo:full:mpropose-ack-pre} is a fast quorum with size $\floor{\frac{r}{2}} + f$.
In this case, we let $\widehat{Q} = Q$ and
$\forall j \in Q \cdot \widehat{t}_j = t_j$, where $t_j$ is defined at \refline{algo:full:mpropose-ack-recv}.
Since $|Q| = \floor{\frac{r}{2}} + f$ and $f \geq 1$, we have $|\widehat{Q}| \geq \floor{\frac{r}{2}} + 1$, as required.
If the computation of $t$ occurs at \refline{algo:full:rec-max},
we have two situations depending on the condition at \refline{algo:full:coordinator-check}.
Let $I$ be the set computed at \refline{algo:full:intersection}, i.e., the intersection between the recovery quorum $Q$ (defined at \refline{algo:full:mrec-ack-recv}) and the fast quorum $\quorums[\id][p]$
($\quorums[\id][p]$ is known by Invariant \ref{inv:not-start-then-payload} and Invariant \ref{inv:not-start-at-mrec-ack}).
First, consider the case in which the condition at \refline{algo:full:coordinator-check} holds.
In this case, we let $\widehat{Q} = Q$ and
$\forall j \in Q \cdot \widehat{t}_j = t_j$, where $t_j$ is defined at \refline{algo:full:mrec-ack-recv}.
Since $|Q| = r - f$ and $f \leq \floor{\frac{r - 1}{2}}$,
we have $|\widehat{Q}| \geq \floor{\frac{r}{2}} + 1$, as required.
Now consider the case in which the condition at \refline{algo:full:coordinator-check} does not hold.
Given that the fast quorum size is $\floor{\frac{r}{2}} + f$ and the size of the recovery quorum $Q$
is $r - f$, we have that $I$ contains at least $(\floor{\frac{r}{2}} + f) - f = \floor{\frac{r}{2}}$ fast quorum processes.
Note that, since $t$ is computed at \refline{algo:full:rec-max}, we have that $\forall k \in Q \cdot \ab_k = 0$ (\refline{algo:full:previously-accepted}).
For this reason, each process in $Q$ had $\bal[\id] = 0$ (\refline{algo:full:mrec-first-see}) when it received the first $\MRec(\id, \_)$ message.
Moreover, as each process in $I$ reported $\RECOVERP$ (i.e., $\forall k \in I \cdot \ph_k = \RECOVERP$ at \refline{algo:full:coordinator-check}),
the $\phase[\id]$ was $\PROPOSE$ (\refline{algo:full:mrec-at-propose}) when the process received the first $\MRec(\id, \_)$ message.
It follows that each of these processes computed their timestamp proposal in the $\MPropose$ handler at \refline{algo:full:mpropose-proposal} (not in the $\MRec$ handler at \refline{algo:full:mrec-proposal}).
Thus, these processes have proposed a timestamp at least as high as the one from the initial coordinator.
In this case, we let $\widehat{Q} = I \union \{ \initial_p(\id) \}$,
$\forall j \in I \cdot \widehat{t}_j = t_j$ where $t_j$ is defined at \refline{algo:full:mrec-ack-recv} and
$\widehat{t}_{\initial_p(\id)}$ be the timestamp sent by $\initial_p(\id)$ in its
$\MProposeAck(\id, \_)$ message.
Since $|I| \geq \floor{\frac{r}{2}}$ and $\initial_p(\id) \not\in I$,
we have $|\widehat{Q}| \geq \floor{\frac{r}{2}} + 1$, as required.
\qed

\bigskip

We now prove that \SYS ensures Ordering.
Consider two commands $c$ and $c'$ submitted during a run of \SYS with identifiers $\id$ and $\id'$.
By \refprop{prop:agreement}, all the processes agree on the final timestamp of a command.
Let $t$ and $t'$ be the final timestamp of $c$ and $c'$, respectively.

\begin{lemmasc}\label{lemma:assym}
  \upshape
  If $c \mapsto_i c'$ then $\tup{t, id} < \tup{t', id'}$.
\end{lemmasc}
\paragraph{\it Proof.}
By contradiction, assume that $c \mapsto_i c'$, but $\tup{t', id'} < \tup{t, id}$.
Consider the point in time when $i$ executes $c$ (\refline{algo:full:execute}).
By Validity, this point in time is unique.
Since $c \mapsto_i c'$, process $i$ cannot have executed $c'$ before this time.
Process $i$ may only execute $c$ once it is in $\mathit{ids}$ (\refline{algo:full:order}).
Hence, $t \leq h[\hspace{1pt}\floor{\frac{r}{2}}\hspace{1pt}]$ (\refline{algo:full:stable-and-committed}).
From \reftheorem{theorem:stable}, $t$ is stable at $i$.
As $\tup{t', id'} < \tup{t, id}$, we get $t' \leq t$, and by \refprop{prop:stability}, $\id' \in \commit \union \execute$ at $i$.
Then, since $t' \leq t \leq h[\hspace{1pt}\floor{\frac{r}{2}}\hspace{1pt}]$ and $c'$ cannot have been executed before $c$ at process $i$, we have $\id' \in \mathit{ids}$.
But then as $\tup{t', id'} < \tup{t, id}$, $c'$ is executed before $c$ at process $i$ (\refline{algo:full:order}). This contradicts $c \mapsto_i c'$.\qed

\begin{lemmasc}\label{lemma:path}
  \upshape
  If $c \mapsto c'$ then whenever a process $i$ executes $c'$, some process has already executed $c$.
\end{lemmasc}
\paragraph{\it Proof.}
Assume that a process $i$ executed $c'$.
By the definition of $\mapsto$, either $c \leadsto c'$, or $c \mapsto_j c'$ at some process $j$.
Assume first that $c \leadsto c'$.
By definition of $\leadsto$, $c$ returns before $c'$ is submitted.
This requires that command $c$ is executed at least at one replica for each of the partition it accesses.
Hence, at the time $c'$ is executed at $i$, command $c$ has already executed elsewhere, as required.

Assume now that $c \mapsto_j c'$ for some process $j$.
Then $c$ and $c'$ access a common partition, say $p$, and by Lemma~\ref{lemma:assym}, $\tup{t, id} < \tup{t', id'}$.
Consider the point in time $\tau$ when $i$ executes $c'$.
Before this, according to \refline{algo:full:stable-recv}, process $i$ receives an $\MStable(\id')$ message from some process $k \in \procs_p$.
Let $\tau' < \tau$ be the moment when $k$ sends this message by executing \refline{algo:full:stable-send}.
According to \refline{algo:full:order}, $\id'$ must belong to $\mathit{ids}$ at time $\tau'$.
Hence, $t' \leq h[\hspace{1pt}\floor{\frac{r}{2}}\hspace{1pt}]$ (\refline{algo:full:stable-and-committed}).
From \reftheorem{theorem:stable}, $t'$ is stable at $k$.
As $\tup{t, id} < \tup{t', id'}$, we have $t \le t'$, and by \refprop{prop:stability}, $\id \in \commit \union \execute$ holds at time $\tau'$ at process $k$.
Since $t \le t' \leq h[\hspace{1pt}\floor{\frac{r}{2}}\hspace{1pt}]$, either $k$ already executed $c$, or $\id \in \mathit{ids}$.
In the latter case, because $\tup{t, id} < \tup{t', id'}$, command $c$ is executed before $k$ sends the $\MStable$ message for $\id'$.
Hence, in both cases, $c$ is executed no later than $\tau' < \tau$, as required.\qed

\paragraph{\it Proof of Ordering.}
By Validity, cycles of size one are prohibited.
By \reflemma{lemma:path}, so are cycles of size two or greater.\qed

\subsection{Proof of Liveness}
\label{sec:correctness-liveness}

We now prove the Liveness property of the PSMR specification.
For simplicity, we assume that links are reliable, i.e., if a message is sent between two correct processes then it is eventually delivered.
In the following, we use $\dom(m)$ to denote the domain of mapping $m$ and $\img(m)$ to denote its image.
We let $\id$ be the identifier of some command $c$, so that $\procs_c$ denotes the set of processes replicating the partitions accessed by $c$.

\begin{lemmasc}
  \label{lemma:after-commit-at-least-one-is-alive}
  \upshape
  Assume that $\id \in \commit \union \execute$ at some process from a partition $p$.
  Then $\id \in \dom(\Attached)$ and $\id \not\in \start$ at some correct process from each partition $q$ accessed by command $c$.
\end{lemmasc}

\paragraph{\it Proof.}

Since $\id \in \commit \union \execute$ at a correct process from $p$, an $\MCommit(\id, \_)$ has been sent by some process from each partition accessed by $c$ (\refline{algo:full:mcommit-recv}).
In particular, it has been sent by some process from partition $q$.
By \refprop{prop:bump-at-majority}, there is a majority of processes $Q$ from partition $q$ that called $\FunTs(\id, \_)$, generating a promise attached to $\id$ (\refline{algo:full:attached}), and thus, have $\id \in \dom(\Attached)$.
Since at most $f$ of these processes can fail, at least some process $j \in Q$ is correct.
Moreover, since $j$ has generated a promise attached to $\id$, it is impossible to have $\id \in \start$ at $j$ (see the $\MPropose$ and $\MRec$ handlers where $\FunTs(\id)$ is called).
Thus $\id \in \dom(\Attached)$ and $\id \not\in \start$ at $j$, as required.
\qed

\begin{lemmasc}
  \label{lemma:not-start-at-one-then-at-all-and-commit}
  \upshape
  Assume that $\id \not\in \start$ at some correct process $i$ from partition $p$.
  Then eventually every correct process $j$ from some partition $q$ accessed by command $c$ has $\id \not\in \start$ and receives an $\MCommit(\id, \_)$ sent by some process from partition $p$.
\end{lemmasc}

\paragraph{\it Proof.}
We consider the case where $\id \not\in \commit \union \execute$ never holds at $j$ (if it does hold, then $j$ has $\id \not\in \start$ and received an $\MCommit(\id, \_)$ sent by some process from partition $p$, as required).

First, assume that $\id \in \commit \union \execute$ eventually holds at process $i$.
By \reflemma{lemma:after-commit-at-least-one-is-alive}, there exists a correct process $k$ from $q$ that has $\id \in \dom(\Attached)$.
Due to \refline{algo:full:mvotes-send}, $k$ continuously sends an $\MPromises$ message to all the processes from $q$, including $j$.
Note that, since $\id \in \dom(\Attached)$ at $k$, this $\MPromises$ message contains a promise attached to $\id$.
Once $j$ receives such a message, since it has $\id \not \in \commit \union \execute$ (\refline{algo:full:attached-not-committed}), it sends an $\MCommitRequest(\id)$ to $\procs_c$ (\refline{algo:full:mcommit-request-send}), and in particular to process $i$.
Since $\id \in \commit \union \execute$ at $i$ (\refline{algo:full:mcommit-request-pre}), process $i$ replies with an $\MPayload(\id, \_, \_)$ and $\MCommit(\id, \_)$.
Once $j$ processes such messages, it has $\id \not\in \start$ and received an $\MCommit(\id, \_)$ by process $i$, a correct process from partition $p$, as required.

Now, assume that $\id \in \pending$ holds forever at process $i$.
Consider the moment $\tau_0$ when the variable $\leader_p$ stabilizes.
Let process $l$ be $\leader_p$.
Assume that $\id \not\in \commit \union \execute$ at processes $i$ and $l$ forever
(the case where it changes to $\id \in \commit \union \execute$ at process $i$ is covered above; the case for process $l$ can be shown analogously).
Due to \refline{algo:full:liveness-pending}, process $i$ sends an $\MPayload(\id, \_, \_)$ message to $\procs_c$ (\refline{algo:full:liveness-mpayload-send}), in particular to $l$ and $j$.
Once $l$ processes this message, it has $\id \in \pending$ forever (since we have assumed that $\id \not\in \commit \union \execute$ at $l$ forever).
Once $j$ processes this message, it has $\id \not\in \start$, as required.
We now prove that eventually process $l$ sends an $\MCommit(\id, \_)$ to $j$.

First, we show by contradiction that the number of $\MRec(\id, \_)$ messages sent by $l$ is finite.
Assume the converse.
After $\tau_0$, due to the check $\leader_p = l$ at \refline{algo:full:liveness-recover-pre} and at \refline{algo:full:mrecnack-pre}, only process $l$ sends $\MRec(\id, \_)$ messages.
Since $\MRec(\id, \_)$ messages by processes other than $l$ are all sent before $\tau_0$, their number is finite.
For this same reason, the number of $\MConsensus(\id, \_, \_)$ messages by processes other than $l$ are also finite.
Thus, each correct process joins only a finite number of ballots that are not owned by $l$.
It follows that the number of $\MRec(\id, \_)$ messages sent by $l$ at \refline{algo:full:liveness-recover} because it joined a ballot owned by other processes (i.e., when $\FunBalLeader(\bal[\id]) \not = l$) is finite. (Note that a single $\MRec(\id, \_)$ can be sent here due to $\bal[\id] = 0$, as process $l$ sets $\bal[\id]$ to some non-zero ballot when processing its first $\MRec(\id, \_)$ message).
For an $\MRec(\id, \_)$ to be sent at \refline{algo:full:mrecnack-recover}, process $l$ has to receive an $\MRecNAck(\id, b)$ with $\bal[\id] < b$ (\refline{algo:full:mrecnack-pre}).
If $\FunBalLeader(b) \not= l$, the number of such $\MRecNAck$ messages is finite as the number of $\MRec(\id, \_)$ and $\MConsensus(\id, \_, \_)$ by processes other than $l$ are finite.
If $\FunBalLeader(b) = l$, the $\MRecNAck(\id, b)$ must be in response to an $\MRec(\id, b)$ or $\MConsensus(\id, \_, b)$ by $l$.
Note that when $l$ sends such a message, it sets $\bal[\id]$ to $b$.
For this reason, process $l$ cannot have $\bal[\id] < b$, and hence this case is impossible.
Thus, there is a point in time $\tau_1 \geq \tau_0$ after which the condition at \refline{algo:full:mrecnack-pre} does not hold at process $l$, and consequently,
process $l$ stops sending new $\MRec(\id, \_)$ messages at \refline{algo:full:mrecnack-recover}, which yields a contradiction.

We have established above that $\id \in \pending$ at process $l$ forever.
We now show that process $l$ sends at least one $\MRec(\id, \_)$ message.
Since $\id \in \pending$ forever, process $l$ executes \refline{algo:full:liveness-recover-pre} for this $\id$ infinitely many times.
If $l$ does not send at least one $\MRec(\id, \_)$ at this line it is because $\bal[\id] > 0$ and $\FunBalLeader(\bal[\id]) = l$ forever.
If so, we have two cases to consider depending on the value of $\bal[\id]$.
In the first case, $\bal[\id] = l$ at $l$ forever. In this case, process $l$ took the slow path by sending to $\procs_p$ an $\MConsensus$ message with ballot $l$ (\refline{algo:full:slow-path}).
We now have two sub-cases.
If any process sends an $\MRecNAck$ to process $l$ (\refline{algo:full:mrecnack-send}), this will make process $l$ send an $\MRec(\id, \_)$ (\refline{algo:full:mrecnack-recover}), as required.
Otherwise, process $l$ will eventually gather $f+1$ $\MConsensusAck$ messages and commit the command.
As we have established that $\id \in \pending$ at process $l$ forever, this sub-case is impossible.
In the second case, we eventually have $\bal[\id] > l$.
Since from some point on, $\FunBalLeader(\bal[\id]) = l$ at $l$, this means that this process sends an $\MRec(\id, \bal[\id])$ (\refline{algo:full:mrec-new-bal}), as required.

We have now established that process $l$ sends a finite non-zero number of $\MRec(\id, \_)$ messages.
Let $b$ be the highest ballot for which process $l$ sends an $\MRec(\id, b)$ message.
We now prove that $l$ eventually sends an $\MCommit(\id, \_)$ to $\procs_c$, in particular to $j$.
Given that at most $f$ processes can fail, there are enough correct processes to eventually satisfy the preconditions of $\MRecAck$ and $\MConsensus$.
First we consider the case where the preconditions of $\MRecAck$ and $\MConsensus$ eventually hold at process $l$.
Since the precondition of $\MRecAck$ eventually holds, process $l$ eventually sends an $\MConsensus(\id, \_, b)$.
Since the precondition of $\MConsensus$ eventually holds, process $l$ eventually sends an $\MCommit(\id, \_)$ to $j$, as required.
Consider now the opposite case where the preconditions of $\MRecAck$ or $\MConsensusAck$ never hold at process $l$.
Since there are enough correct processes to eventually satisfy these preconditions, the fact that they never hold at process $l$ implies that there is some correct process $j$ with $\bal[\id] > b$ (otherwise $j$ would eventually reply to process $l$).
Thus, the precondition of $\MRecNAck$ holds at $j$ (\refline{algo:full:mconsensus-or-mrec-pre-high-ballot}), which causes $j$ to send an $\MRecNAck(\id, b')$ with $b' = \bal[\id] > b$ to $l$.
When $l$ receives such a message, it sends a new $\MRec(\id, b'')$ with some ballot $b'' > b'$ (\refline{algo:full:mrecnack-recover}).
It follows that $b'' > b' > b$, which contradicts the fact that $b$ is the highest ballot sent by $l$, and hence this case is impossible.
\qed

\begin{lemmasc}
  \label{lemma:commit-liveness}
  \upshape
  Assume that $\id \not\in \start$ at some correct process $i$ from some partition $p$ accessed by command $c$.
  Then eventually $\id \in \commit \union \execute$ at every correct process $j \in \procs_c$.
\end{lemmasc}

\paragraph{\it Proof.}
For $c$ to be committed at process $j$, $j$ has to have $\id \not\in \start$ and to receive an $\MCommit(\id, \_)$ from each of the partitions accessed by $c$ (\refline{algo:full:mcommit-recv}).
We prove that process $j$ eventually receives such a message from each of these partitions.
To this end, fix one such partition $q$.
By \reflemma{lemma:not-start-at-one-then-at-all-and-commit}, it is enough to prove that some correct process from partition $q$ eventually has $\id \not\in \start$.
By contradiction, assume that all the correct processes from partition $q$ have $\id \in \start$ forever.
We have two scenarios to consider.
First, consider the scenario where $p = q$.
But this contradicts the fact that $\id \not\in \start$ at process $i$.
Now, consider the scenario where $p \not= q$.
In this scenario we consider two sub-cases.
In the first case, eventually $\id \in \commit \union \execute$ at process $i$.
By \reflemma{lemma:after-commit-at-least-one-is-alive},
there is a correct process from each partition accessed by $c$, in particular from partition $q$, that has $\id \not\in \start$, which contradicts our assumption.
In the second case, $\id \in \pending$ at process $i$ forever
(the case where it changes to $\id \in \commit \union \execute$ is covered above).
Due to \refline{algo:full:liveness-mpayload-send}, process $i$ periodically sends an $\MPayload$ message to the processes in partition $q$.
Once this message is processed, the correct processes in partition $q$ will have $\id \in \payload$.
Since partition $q$ contains at least one correct process, this also contradicts our assumption.
\qed

\begin{definition}
  \label{def:local-votes}
  \upshape
  We define the set of proposals issued by some process $i \in \procs_p$ as $\LocalPromises_i = \img(\Attached) \union \Detached$.
\end{definition}

\begin{lemmasc}
  \label{lemma:no-holes}
  \upshape
  For each process $i \in \procs_p$
  we have that
  $\tup{i,t} \in \LocalPromises_i \implies (\forall u \in \{1,\dots, t \} \cdot \tup{i, u} \in \LocalPromises_i)$.
\end{lemmasc}

\paragraph{\it Proof.}
Follows trivially from \refalgo{algo:full}.
\qed\\

\begin{lemmasc}
  \label{lemma:eventually-stable}
  \upshape
  Consider a command $c$ with an identifier $\id$ and the final timestamp $t$, and assume that $\id \not\in \start$ at some correct process in $\procs_c$.
  Then at every correct process in $\procs_c$, eventually variable $\Promises$ contains all the promises up to $t$ by some set of processes $C$ with $\setsize{C} \geq \floor{\frac{r}{2}} + 1$.
\end{lemmasc}

\paragraph{\it Proof.}
By \reflemma{lemma:commit-liveness}, eventually $\id \in \commit \union \execute$ at all the correct processes in $\procs_c$.
Consider a point in time $\tau_0$ when this happens and fix a process $i \in \procs_c$ from a partition $p$ that has $\id \in \commit$.
Let $C$ be the set of correct processes from partition $p$,
$\MPromises(D^0_j, A^0_j)$ be the $\MPromises$ message sent by each process $j \in C$ in the next invocation of \refline{algo:full:mvotes-send} after $\tau_0$, and $\mathit{ids} = \bigunion \{\dom(A^0_j) \mid j \in C\}$.
Due to \refline{algo:full:mcommit-bump}, by \reflemma{lemma:no-holes} we have that
\begin{quote}
  (*) $\forall j \in C, u \in \{1,\dots,t\} \cdot \tup{j, u} \in \img(A^0_j) \union D^0_j$.
\end{quote}
Note that, for each $\id' \in \mathit{ids}$, we have $\id' \not\in \start$ at some correct process (in particular, at the processes $j$ that sent such $\id'$ in their $\MPromises(D^0_j, A^0_j)$ messages).
Thus, by \reflemma{lemma:commit-liveness}, there exists $\tau_1 \geq \tau_0$ at which $\mathit{ids} \subseteq \commit \union \execute$ at process $i$.
Let $\MPromises(D^1_j, A^1_j)$ be the $\MPromises$ message sent by each process $j \in C$ in the next invocation of \refline{algo:full:mvotes-send} after $\tau_1$.
Since at process $i$ we have that $\mathit{ids} \subseteq \commit \union \execute$, once all these $\MPromises$ messages are processed by process $i$, for each $j \in C$ we have that $\bigunion \{A^1_j[\id'] \mid \id' \in \mathit{ids} \} \subseteq \Promises$ and $D^1_j \subseteq \Promises$ at process $i$.
Given the definition of $\mathit{ids}$ and since $A^0_j \subseteq A^1_j$ and $D^0_j \subseteq D^1_j$,
we also have that $\img(A^0_j) \subseteq \Promises$ and $D^0_j \subseteq \Promises$ at process $i$.
From (*), it follows that $\forall j \in C, u \in \{1,\dots,t\} \cdot \tup{j, u} \in \Promises$ at process $i$.
\qed

\begin{lemmasc}
  \label{lemma:eventually-stable-execute}
  \upshape
  Consider a command $c$ with an identifier $\id$, and assume that $\id \not\in \start$ at some correct process in $\procs_c$.
  Then every correct process $i \in \procs_c$ eventually executes $c$.
\end{lemmasc}

\paragraph{\it Proof.}
Consider a command $c$ with an identifier $\id$, and assume that $\id \not\in \start$ at some correct process in $\procs_c$.
By contradiction, assume further that some correct process $i \in \procs_c$ never executes $c$.
Let $c_0 = c$, $\id_0 = \id$, and $t_0$ be the timestamp assigned to $c$.
Then by \reflemma{lemma:eventually-stable}, eventually in every invocation of the periodic handler at line~\ref{algo:full:executehandler}, we have $h \le t_0$.
By \reflemma{lemma:commit-liveness} and since $i$ never executes $c_0$, eventually $i$ has $\id_0 \in \commit$.
Hence, either $i$ never executes another command $c_1$ preceding $c_0$ in $\mathit{ids}$, or $i$ never receives an $\MStable(\id_0)$ message from some correct process $j$ eventually indicated by $\procs_{c_0}^i$.
The latter case can only be due to the loop in the handler at line~\ref{algo:full:executehandler} being stuck at an earlier command $c_1$ preceding $c_0$ in $\mathit{ids}$ at $j$.
Hence, in both cases there exists a command $c_1$ with an identifier $\id_1$ and a final timestamp $t_1$ such that $\tup{t_1, \id_1} < \tup{t_0, \id_0}$ and some correct process $i_1 \in \procs_{c_1}$ never executes $c_1$ despite eventually having $\id_1 \in \commit$.
Continuing the above reasoning, we obtain an infinite sequence of commands $c_0, c_1, c_2, \ldots$ with decreasing timestamp-identifier pairs such that each of these commands is never executed by some correct process. But such a sequence cannot exist because the set of timestamp-identifier pairs is well-founded. This contradiction shows the required.\qed

\paragraph{\it Proof of Liveness.}
Assume that some command $c$ with identifier $\id$ is submitted by a correct process or executed at some process.
We now prove that it is eventually executed at all correct processes in $\procs_c$.
By \reflemma{lemma:eventually-stable} and \reflemma{lemma:eventually-stable-execute}, it is enough to prove that eventually some correct process in $\procs_c$ has $\id \not\in \start$.
First, assume that $\id$ is submitted by a correct process $i$.
Due to the precondition at \refline{algo:full:submit-pre}, $i \in \procs_c$.
Then $\phase[\id]$ is set to $\PROPOSE$ at process $i$ when $i$ sends the initial $\MPropose$ message, and thus $\id \not\in \start$ as required.
Assume now that $\id$ is executed at some (potentially faulty) process.
By \reflemma{lemma:after-commit-at-least-one-is-alive}, there exists some correct process in $\procs_c$ with $\id \not\in \start$, as required.
\qed

\section{Pathological Scenarios with Caesar and EPaxos}
\label{sec:app_issues}

Let us consider 3 processes: \A, \B, \C.
Assume that all the commands are conflicting.
Assume further, that if a process proposes a command $x$, then this command is proposed with timestamp $x$.
In Caesar, timestamps proposed must be unique.
In the example below, this is ensured by assigning timestamps to processes round-robin.
Further details about Caesar are provided in \refsec{sec:sys_comparison}.
The (infinite) example we consider is as follows:
\begin{itemize}
  \item \A proposes $1$, $4$, $7$, ...
  \item \B proposes $2$, $5$, $8$, ...
  \item \C proposes $3$, $6$, $9$, ...
\end{itemize}

First, \A proposes $1$.
When command $1$ reaches \B, \B has already proposed $2$, and thus command $1$ is blocked on it.
When $2$ reaches \C, \C has already proposed $3$, and thus command $2$ is blocked on it.
This keeps going forever, as depicted in the diagram below (where $\leftarrow$ denotes "blocked on").

\begin{displaymath}
  \begin{array}{lllll}
    \A: & 1              & 4 \leftarrow 3 & 7 \leftarrow 6 & \ldots \\
    \B: & 2 \leftarrow 1 & 5 \leftarrow 4 & 8 \leftarrow 7 & \ldots \\
    \C: & 3 \leftarrow 2 & 6 \leftarrow 5 & 9 \leftarrow 8 & \ldots
  \end{array}
\end{displaymath}

As a result, no command is ever committed. This liveness issue is due to Caesar's wait condition~\cite{caesar}.

In EPaxos, the command arrival order from above results in the following dependency sets being committed:
\begin{itemize}
  \item $\dep[1] = \{\textbf{2}\}$
  \item $\dep[2] = \{\textbf{3}\}$
  \item $\dep[3] = \{1, \textbf{4}\}$
  \item $\dep[4] = \{1, 2, \textbf{5}\}$
  \item $\dep[5] = \{2, 3, \textbf{6}\}$
  \item $\dep[6] = \{1, 3, 4, \textbf{7}\}$
  \item \dots
\end{itemize}
These committed dependencies form a strongly connected component of unbounded size
\cite{epaxos, leaderless}.
As result, commands are never executed.

\fi

\end{document}